\documentclass[twocolumn,english,prb,aps,showpacs,superscriptaddress]{revtex4}
\usepackage[T1]{fontenc}
\usepackage[latin9]{inputenc}
\usepackage{graphicx}

\makeatletter

\usepackage{babel}

\begin{document}

\title{Gap nodes induced by coexistence with antiferromagnetism in iron-based
superconductors}

\author{S. Maiti}

\affiliation{Department of Physics, University of Wisconsin, Madison, Wisconsin
53706, USA}

\author{R. M. Fernandes}

\affiliation{Department of Physics, Columbia University, New York,
New York 10027, USA}
\affiliation{Theoretical Division, Los Alamos
National Laboratory, Los Alamos, NM, 87545, USA}

\author{A.V. Chubukov}

\affiliation{Department of Physics, University of Wisconsin, Madison, Wisconsin
53706, USA}

\date{\today}

\begin{abstract}
We investigate the pairing in iron pnictides in the coexistence phase,
which displays both superconducting and antiferromagnetic orders.
By solving the pairing problem on the Fermi surface reconstructed
by long-range magnetic order, we find that the pairing interaction
necessarily becomes angle-dependent, even if it was isotropic in the
paramagnetic phase, which results in an angular variation of the superconducting
gap along the Fermi surfaces. We find that the gap has no nodes for
a small antiferromagnetic order parameter $M$, but may develop accidental
nodes for intermediate values of $M$, when one pair of the reconstructed
Fermi surface pockets disappear. For even larger $M$, when the other
pair of reconstructed Fermi pockets is gapped by long-range magnetic
order, superconductivity still exists, but the quasiparticle spectrum
becomes nodeless again. We also show that the application of an external
magnetic field facilitates the formation of nodes. We argue that this
mechanism for a nodeless-nodal-nodeless transition explains recent
thermal conductivity measurements of hole-doped (Ba$_{1-x}$K$_{x}$)Fe$_{2}$As$_{2}$
{[}J-Ph. Read \textit{et al} arXiv:1105.2232].
\end{abstract}
\maketitle

\section{Introduction}

Exploring the different regions in the phase diagram of iron based
superconductors (FeSCs) is an important step towards developing a
unified understanding of the physics of superconductivity in these
systems. A typical phase diagram of a FeSC in the $(T,x)$ plane,
where $x$ is doping, shows a metallic antiferromagnetic order, also
called spin density wave (SDW), below $T_{N}$ at $x=0$. Upon doping,
the SDW order parameter is suppressed and superconductivity (SC) emerges
with maximum $T_{c}(x)$ near the point where $T_{c}(x)$ exceeds
$T_{N}(x)$ (an \textquotedbl{}optimal doping\textquotedbl{}).
The gap symmetry at optimal doping is most likely $s^{+-}$, but the
structure of the s-wave gap varies from one material to another: the
gap is nodeless in Ba(Fe$_{1-x}$Co$_{x}$)$_{2}$As$_{2}$ \cite{BFCA_1,BFCA_2,BFCA_3}
and (Ba$_{1-x}$K$_{x}$)Fe$_{2}$As$_{2}$ \cite{BKFA_1,BKFA_2,BKFA_3},
but has nodes in BaFe$_{2}$(As$_{1-x}$P$_{x}$)$_{2}$ \cite{BFAP_1,BFAP_2}.

In this article we investigate the behavior of the $s^{+-}$ SC gap
in the underdoped regime of the Ba(Fe$_{1-x}$Co$_{x}$)$_{2}$As$_{2}$
and (Ba$_{1-x}$K$_{x}$)Fe$_{2}$As$_{2}$ systems, where $T_{c}(x)<T_{N}(x)$,
and superconductivity emerges in a continuous fashion from a pre-existing
SDW order. The microscopic coexistence of SC and SDW below $T_{c}(x)$
has been reported by magnetization and NMR experiments \cite{coexist_1,coexist_2,coexist_3,coexist_4}.
\begin{figure}
\begin{centering}
\includegraphics[width=0.9\columnwidth]{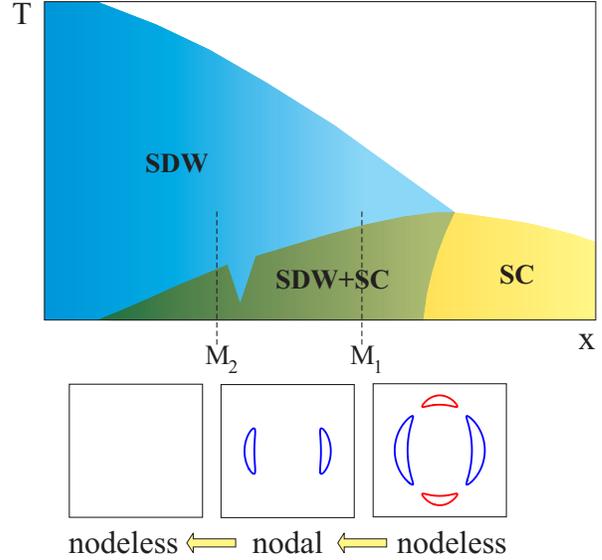}
\par\end{centering}

\caption{Schematic phase diagram showing the nodeless-nodal-nodeless transitions
inside the region where superconductivity (SC) and spin density wave
(SDW) coexist. Each transition is roughly associated with the disappearance
of one pair of magnetically reconstructed Fermi surface pockets. $M_{1}$
and $M_{2}$ refer to the values of the SDW order parameter $M$ at
which the two pairs of pockets disappear respectively. A dip in $T_{c}$
is observed near $M_{2}$, when both pockets disappear. \label{fig_phase_diagram} }
\end{figure}

The electronic structure of FeSCs in the paramagnetic phase consists
of near-circular hole pockets and elliptical electron pockets. We
assume, following earlier works~\cite{Andrey,parker_09,rafael,vvc_10,fernandes_optics,eremin_09,fernandes_11},
that the interaction leading to SDW order is independent on the angle
along the FS~\cite{comm}. In this situation, the Brillouin Zone
(BZ) is reduced once the SDW order sets in, and the Fermi surface
(FS) is reconstructed into four banana-like pockets (right panel in
Fig. \ref{fig_phase_diagram}). As the SDW order parameter $M$ increases,
the reconstructed pockets shrink and eventually disappear. This happens
in two stages: first, at $M=M_{1}$, the number of pockets shrinks
from four to two, (middle panel in Fig. \ref{fig_phase_diagram}),
and then the remaining two pockets vanish at a larger value $M=M_{2}$
(left panel in Fig. \ref{fig_phase_diagram}). Because these reconstructed
FSs are formed by mixing electron and hole bands, on which the $s^{+-}$
SC order parameter has different signs in the absence of SDW order,
one could naively expect that the SC gap on the reconstructed FSs
alternates between plus and minus signs, implying that it must have
nodes. However, calculations show \cite{parker_09} that this is not
the case because SDW order, with e.g. $\left\langle S_{x}\right\rangle \neq0$,
mixes hole and electron FSs with opposite $\sigma_{z}$ components
of electron spins. Because a spin-singlet s-wave gap changes sign
under $\sigma_{z}\to-\sigma_{z}$, the sign change imposed by the
$s^{+-}$ gap structure is compensated by the sign change due to the
flip of $\sigma_{z}$, and, as a result, the gaps on the reconstructed
FSs retain the same nodeless structure as in the absence of magnetic
 order.

Experiments near the onset of the co-existence phase, where $M$ is
small, are in agreement with this reasoning. For instance, the ratio
$\kappa/T$, where $\kappa$ is the in-plane thermal conductivity,
tends to zero in the $T\to0$ limit, as it is expected for a superconductor
with a nodeless gap. However, recent thermal conductivity measurements
on (Ba$_{1-x}$K$_{x}$)Fe$_{2}$As$_{2}$ deep in the coexistence
phase \cite{JPR} found that there is a range of dopings where $\kappa/T$
remains finite at $T=0$, as it happens in a superconductor with line
nodes. At even smaller dopings, $\kappa/T$ again vanishes at $T=0$.
Application of a magnetic field makes the doping dependence more smooth
and extends the range of nodal behavior.

In this article we explain the sequence of nodeless-nodal-nodeless
behavior as a transition from a nodeless $s^{+-}$ gap in the
region where all four reconstructed FS pockets are present, to an
$s^{+-}$ gap with accidental nodes in the region where two FS
pockets are present, to the fully gapped quasiparticle spectrum in
the region where the remaining two reconstructed FS pockets are
also gapped (see Fig. \ref{fig_phase_diagram}). We show that the
pairing interaction in the coexistence phase acquires an
additional angular dependence as it gets dressed by
angle-dependent coherence factors associated with the SDW order.
The angular dependence of the interactions in turn gives rise to
an angular dependence of the superconducting gap. This effect is
weak in the limit of small $M$, considered in Ref.
\onlinecite{parker_09}, when all four reconstructed Fermi pockets
are present, but is strong and gives rise to gap nodes when $M$
becomes large enough to gap out one pair of Fermi pockets. At even
larger $M$, deeper in the coexistence region, the remaining pair
of Fermi pockets is gapped out. In this situation,
superconductivity still develops over some range of
dopings~\cite{vvc_10}, but the quasiparticle excitation spectrum
remains gapped irrespective of the structure of the SC gap, and
$\kappa/T$ again vanishes at $T=0$. Concurrently, the
superconducting transition temperature $T_{c}$ shows a minimum
when the remaining pockets vanish. The theoretical behavior across
the SDW-SC coexistence region is shown schematically in Fig.
\ref{fig_phase_diagram}. It agrees well with the experimental
results of Refs. \onlinecite{JPR,ruslan_private.}
%SM commented
%SM re-inserted after APS

The paper is organized as follows. In Sec. \ref{sec:approach} we
describe the formalism listing out the details of the FS geometries,
the nature of the reconstructed FSs, and the generic form of the SC
gap in the coexistence region. In Sec. \ref{sec:results} we discuss
the solutions for the gap obtained within this formalism, show when
and how nodes arise in the coexistence region, and provide analytical
explanation of the results. We also briefly discuss the role played
by bands that do not participate in the formation of SDW. In Sec.\ref{sec:magnetic field}
we show that the doping range where the SC gap has nodes is enhanced
by applying a magnetic field. We state our final conclusions in Sec.
\ref{sec:conclusion}.

\section{The pairing problem on the reconstructed Fermi surface}

\label{sec:approach}

Our point of departure is a microscopic band model of interacting
fermions located near hole or electron pockets. The electronic structure
of FeSCs consists of two or three hole pockets, centered at $(0,0)$
in the folded zone (two Fe atoms per unit cell), and two hybridized
electron pockets centered at $(\pi,\pi)$. The interactions between
low-energy fermions are generally angle-dependent already in the normal
state, before either SDW or SC order sets in, which gives rise to
angle-dependent SDW and SC gaps even outside the coexistence region.

As we said, we follow earlier works and consider a simplified model
in which we neglect the angular dependencies of the interactions in
the normal state, and approximate the normal-state interactions as
constants \cite{Andrey,parker_09,rafael,vvc_10,fernandes_optics,eremin_09,fernandes_11}.
We further neglect the hybridization and perform calculations in the
unfolded zone (one Fe atom per unit cell), in which electron pockets
are ellipses centered at $(0,\pi)$ and $(\pi,0)$ and hole pockets
are at $(0,0)$ and $(\pi,\pi)$. Finally, we neglect the third hole
pocket and only consider two circular hole pockets centered at $(0,0)$
in the unfolded zone (in orbital notations these are $d_{xz}/d_{yz}$
pockets). Earlier works have found \cite{eremin_09} that SDW magnetic
order emerges in this model with ordering vector $\mathbf{Q}=(0,\pi)$
or $\mathbf{Q}=(\pi,0)$, mixing one hole and one electron pocket.
For definiteness we set $\mathbf{Q}=(0,\pi)$ in which case one hole
pocket at $(0,0)$ and one electron pocket at $(0,\pi)$ are mixed
up. The other two pockets are not participating in SDW and remain
intact once the SDW order sets in. In line with our goal, we first
consider only the 2-pocket model with one circular hole and one elliptical
electron pocket, which get reconstructed in the SDW phase. We discuss
the role of the other pockets later in the next section.

To quadratic order in the fermions, the Hamiltonian of the 2-pocket
model in the SDW phase is $H_{0}+H_{SDW}$, where

\begin{eqnarray}
H_{0} & = & \sum_{\mathbf{k},\sigma}\left(\varepsilon_{\mathbf{k}}^{c}c_{\mathbf{k}\sigma}^{\dagger}c_{\mathbf{k}\sigma}+\varepsilon_{\mathbf{k}+\mathbf{Q}}^{f}f_{\mathbf{k}+\mathbf{Q}\sigma}^{\dagger}f_{\mathbf{k}+\mathbf{Q}\sigma}^{}\right)\nonumber \\
H_{SDW} & = & \sum_{\mathbf{k}\sigma}M\left(\sigma c_{\mathbf{k}\sigma}^{\dagger}f_{\mathbf{k}+\mathbf{Q}\sigma}^{}+\mathrm{h.c.}\right)\label{Hamiltonian_2_start}\end{eqnarray}

\begin{figure*}[htp]
 \includegraphics[width=0.6\columnwidth]{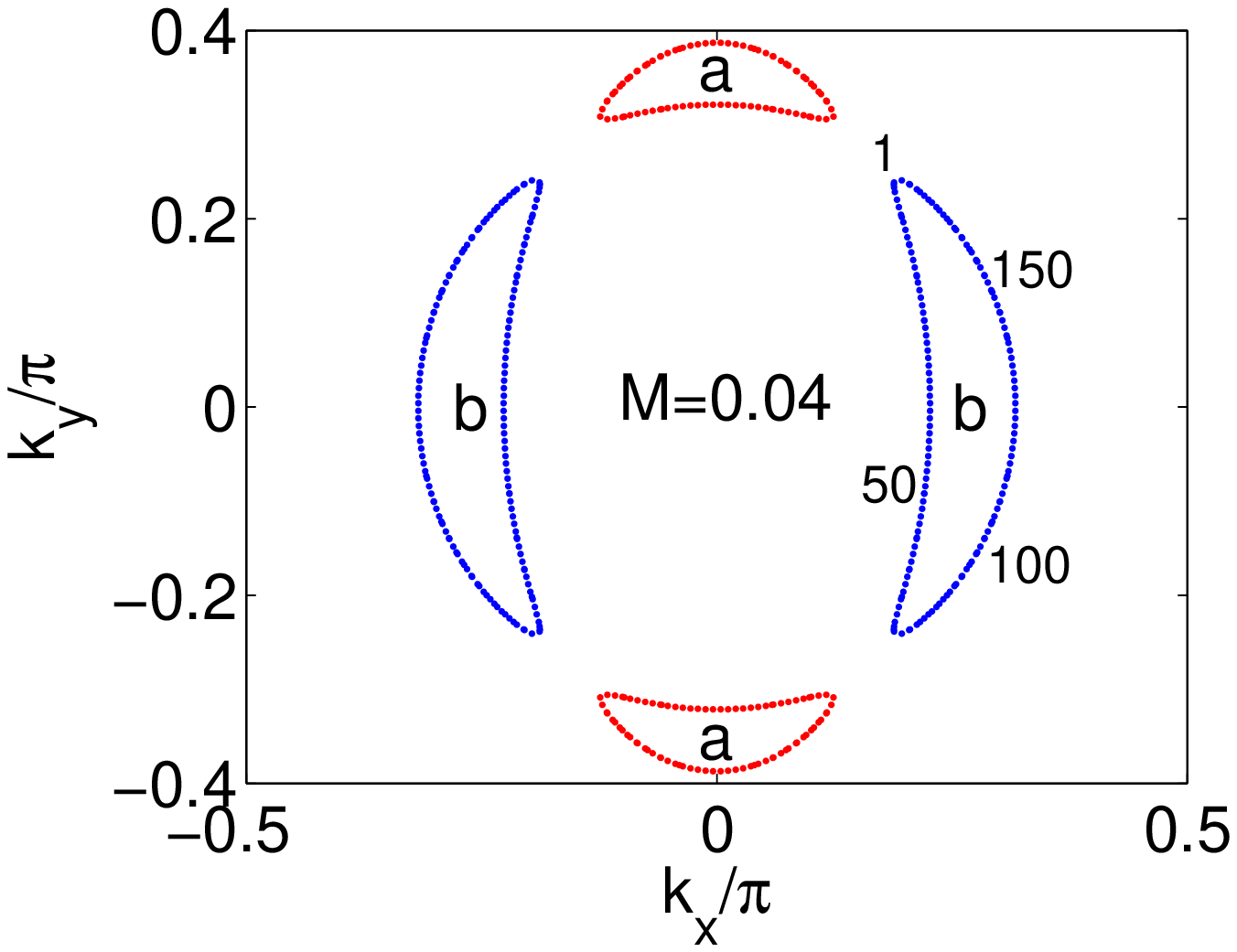}
 \includegraphics[width=0.6\columnwidth]{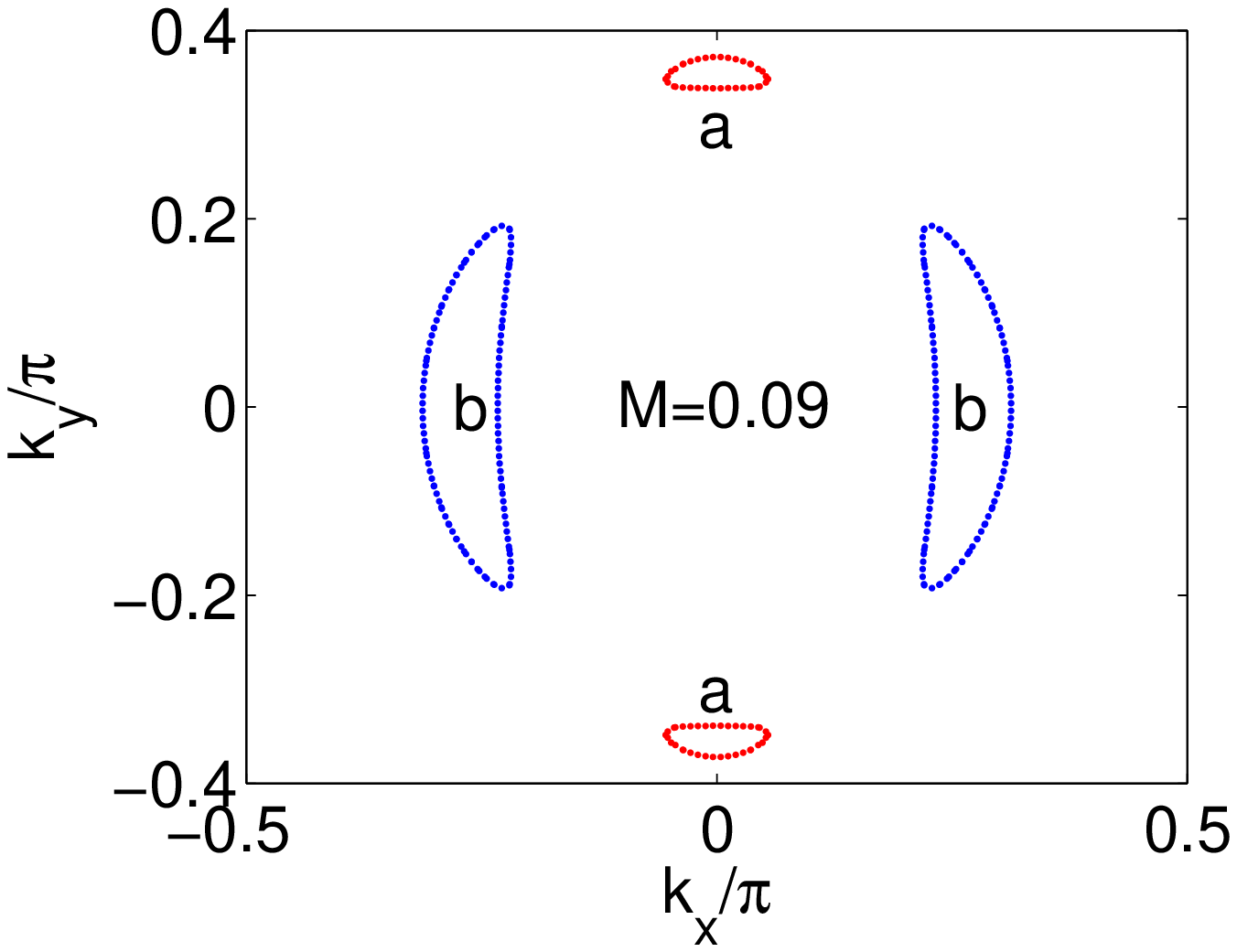}
\includegraphics[width=0.6\columnwidth]{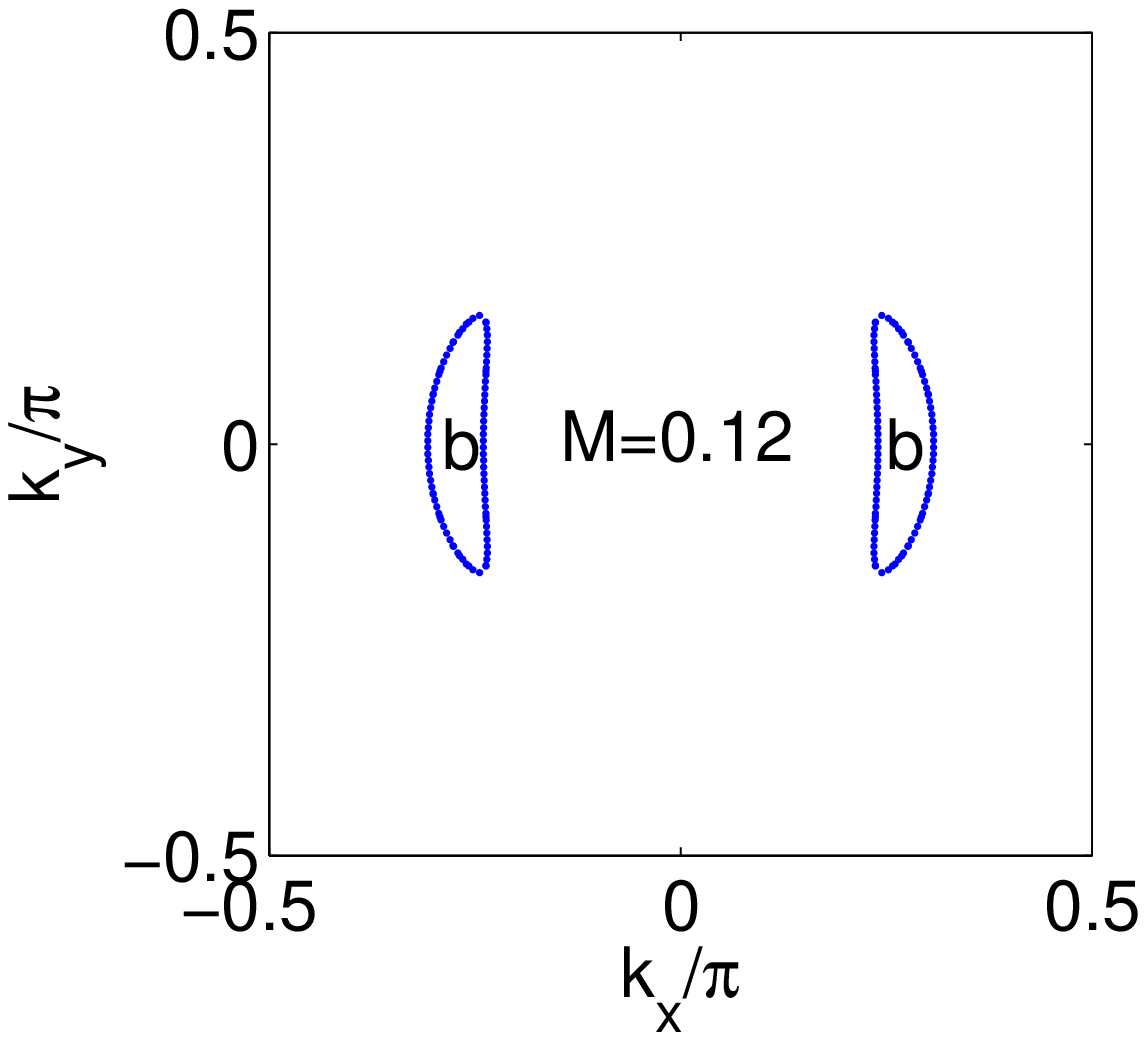}

\caption{\label{fig:FS} Reconstructed $a$ and $b$ Fermi pockets for several
values of the SDW order parameter $M$. The $a$-pockets are smaller
and disappear first upon increasing $M$. At even larger $M$, the
$b$-pockets also disappear (not shown). Each pocket was discretized
into a finite number of points indicated in the left figure for one
of the $b$ pockets. Our convention is such that the numbering for
each pocket starts at the tip of a {}``banana'' and goes first along
the {}``inner'' side of the pocket and then along the {}``outer''
side.}

\end{figure*}

Here $c$ and $f$ are fermionic operators for hole and electron states,
respectively, $\sigma$ is $\pm1$, and $M$ is the SDW order parameter,
which we treat below as a variable. In practice, larger $M$ correspond
to smaller dopings, whereas smaller $M$ correspond to doping close
to the optimal one. The fermionic dispersions are

\begin{eqnarray}
\varepsilon_{\mathbf{k}}^{c} & = & \mu_{c}-\frac{k_{x}^{2}+k_{y}^{2}}{2m}\nonumber \\
\varepsilon_{\mathbf{k}+\mathbf{Q}}^{f} & = & -\mu_{f}+\frac{k_{x}^{2}}{2m_{x}}+\frac{k_{y}^{2}}{2m_{y}}\end{eqnarray}
 where $m$, $m_{x}$, $m_{y}$ are the effective band masses of the
fermions. For definiteness, we set the interatomic spacing to one
and use $\mu_{c}=\mu_{f}=\mu$, $m=1/(2\mu)$, $m_{x}=0.5/(2\mu)$,
and $m_{y}=1.5/(2\mu)$. These parameters are chosen to give the FS
geometry as in Fig. \ref{fig:FS}. We will also measure $M$ in units
of $2\mu$.

The $M$ term in (\ref{Hamiltonian_2_start}) couples $c$ and $f$
operators such that the eigenstates of (\ref{Hamiltonian_2_start})
are coherent superpositions of electrons and holes. These are described
by new fermionic operators $a$ and $b$, with dispersion $E_{\mathbf{k}}^{a,b}$
that vanishes at the reconstructed FSs. The transformation to the
new operators is

\begin{eqnarray}
\left(\begin{array}{c}
\sigma a_{\mathbf{k}\sigma}\\
b_{\mathbf{k}\sigma}\end{array}\right) & = & \left(\begin{array}{cc}
-\sigma\cos\theta_{\mathbf{k}} & \sin\theta_{\mathbf{k}}\\
-\sin\theta_{\mathbf{k}} & -\sigma\cos\theta_{\mathbf{k}}\end{array}\right)\left(\begin{array}{c}
c_{\mathbf{k}\sigma}\\
f_{\mathbf{k+Q}\sigma}\end{array}\right)\label{transformation}\end{eqnarray}
 where

\begin{eqnarray}
\cos\theta_{\mathbf{k}} & = & \frac{M}{\sqrt{M^{2}+(E^{-})^{2}}}\nonumber \\
\sin\theta_{\mathbf{k}} & = & \frac{E^{-}}{\sqrt{M^{2}+(E^{-})^{2}}}\label{dec_7_1}\end{eqnarray}
 and

\begin{equation}
E^{-}=\frac{\varepsilon^{c}-\varepsilon^{f}}{2}-\sqrt{\left(\frac{\varepsilon^{c}-\varepsilon^{f}}{2}\right)^{2}+M^{2}}\label{dec_7_2}\end{equation}
 such that

\begin{eqnarray}
\cos2\theta_{\mathbf{k}} & = & \frac{\frac{\varepsilon^{c}-\varepsilon^{f}}{2}}{\sqrt{M^{2}+\left(\frac{\varepsilon^{c}-\varepsilon^{f}}{2}\right)^{2}}}\nonumber \\
\sin2\theta_{\mathbf{k}} & = & -\frac{M}{\sqrt{M^{2}+\left(\frac{\varepsilon^{c}-\varepsilon^{f}}{2}\right)^{2}}}\label{dec_7_3}\end{eqnarray}

In terms of the new operators, \begin{equation}
H_{0}+H_{SDW}=\sum_{\mathbf{k},\sigma}\left(E_{\mathbf{k}}^{a}a_{\mathbf{k}\sigma}^{\dagger}a_{\mathbf{k}\sigma}+E_{\mathbf{k}}^{b}b_{\mathbf{k}\sigma}^{\dagger}b_{\mathbf{k}\sigma}\right)\label{Hamiltonian_new}\end{equation}
 where

\begin{equation}
E_{\mathbf{k}}^{a,b}=\frac{\varepsilon^{c}+\varepsilon^{f}}{2}\pm\sqrt{\left(\frac{\varepsilon^{c}-\varepsilon^{f}}{2}\right)^{2}+M^{2}}\end{equation}
 are the quasi-particle excitation energies of the SDW state (the
plus sign corresponds to $a$ fermions, the minus sign corresponds
to $b$ fermions).

The condition $E_{\mathbf{k}}^{a,b}=0$ defines the new reconstructed
FSs. For small enough $M$, there are two pairs of banana-shaped reconstructed
pockets, the $a$-pockets and the $b$-pockets (see Fig. \ref{fig:FS}),
with the latter larger than the former. At the critical $M=M_{1}$
given by

\begin{equation}
M_{1}=\frac{1}{4}\frac{|m-m_{y}|}{\sqrt{mm_{y}}}\label{M_1}\end{equation}
 the \textbf{$a$}-pockets disappear while the $b$-pockets remain
(Fig. \ref{fig:FS}-right). At even larger $M=M_{2}$ given by

\begin{equation}
M_{2}=\frac{1}{4}\frac{|m-m_{x}|}{\sqrt{mm_{x}}}\label{M_2}\end{equation}
 the $b$-pockets also disappear, i.e. all electronic states become
gapped by SDW \cite{comm}. For our set of parameters $M_{1}\approx0.102$
and $M_{2}\approx0.177$.

We now turn to the issue of the pairing on the reconstructed FS. There
are four residual interactions between the original fermions located
near the hole and electron FSs~ \cite{Andrey}: the interaction between
fermionic densities near hole and electron pockets $\left[U_{1}c_{\sigma}^{\dagger}c_{\sigma}f_{\sigma^{\prime}}^{\dagger}f_{\sigma^{\prime}}\right]$,
the exchange interaction between hole and electron pockets $\left[U_{2}c_{\sigma}^{\dagger}f_{\sigma}f_{\sigma^{\prime}}^{\dagger}c_{\sigma^{\prime}}\right]$,
the umklapp pair-hopping interaction in which two fermions near a
hole pocket are converted into two fermions near an electron pocket
and vice-versa $\left[U_{3}/2(c_{\sigma}^{\dagger}f_{\sigma}c_{\sigma^{\prime}}^{\dagger}f_{\sigma^{\prime}}+f_{\sigma}^{\dagger}c_{\sigma}f_{\sigma^{\prime}}^{\dagger}c_{\sigma^{\prime}})\right]$,
and the density-density interaction within each pocket $\left[U_{4}/2(c_{\sigma}^{\dagger}c_{\sigma}c_{\sigma^{\prime}}^{\dagger}c_{\sigma^{\prime}}+f_{\sigma}^{\dagger}f_{\sigma}f_{\sigma^{\prime}}^{\dagger}f_{\sigma^{\prime}})\right]$.
In the absence of SDW order, only $U_{3}$ and $U_{4}$ contribute
to the pairing channel ($U_{3}$ must be larger than $U_{4}$ for
$s^{+-}$ pairing). In the SDW phase, $c$ and $f$ operators are
mixed up, and all four interactions contribute to pairing vertices
for $a$ and $b$ fermions. The interactions $U_{2}$ and $U_{4}$
do not give rise to new physics and only renormalize the values of
the pairing vertices obtained from the $U_{1}$ and $U_{3}$ terms.
To shorten the formulas, we neglect the $U_{2}$ and $U_{4}$ terms
and approximate $H_{int}$ by

\begin{eqnarray}
H_{int} & = & \sum_{[1234]}U_{1}\, c_{1\sigma}^{\dagger}f_{2\sigma'}^{\dagger}f_{3\sigma'}c_{4\sigma}\nonumber \\
 &  & +\frac{U_{3}}{2}\left(c_{1\sigma}^{\dagger}c_{2\sigma'}^{\dagger}f_{3\sigma'}f_{4\sigma}+\mathrm{h.c.}\right)\label{1}\end{eqnarray}
 where $\sum_{[1234]}$ denotes the sum over all momenta subject to
$\mathbf{k}_{1}+\mathbf{k}_{2}=\mathbf{k}_{3}+\mathbf{k}_{4}$, and
the summation over repeated spin indices is implied. Converting to
new operators via Eq. (\ref{transformation}) and keeping only the
interactions which contribute to the pairing, we obtain from Eq. (\ref{1})

\begin{eqnarray}
 &  & H_{pair}=\nonumber \\
 &  & \sum_{\mathbf{q},\mathbf{k},\sigma,\sigma'}U_{\mathbf{q},\mathbf{k}}^{aa}\left(a_{\mathbf{q}\sigma}^{\dagger}a_{-\mathbf{q}\sigma'}^{\dagger}a_{-\mathbf{k}\sigma'}a_{\mathbf{k}\sigma}+b_{\mathbf{q}\sigma}^{\dagger}b_{-\mathbf{q}\sigma'}^{\dagger}b_{-\mathbf{k}\sigma'}b_{\mathbf{k}\sigma}\right)\nonumber \\
 &  & +\sum_{\mathbf{q},\mathbf{k},\sigma,\sigma'}U_{\mathbf{q},\mathbf{k}}^{ab}\left(a_{\mathbf{q}\sigma}^{\dagger}a_{-\mathbf{q}\sigma'}^{\dagger}b_{-\mathbf{k}\sigma}b_{\mathbf{k},\sigma}+\mathrm{h.c.}\right)\label{H_pair}\end{eqnarray}

where \begin{eqnarray*}
U_{\mathbf{q},\mathbf{k}}^{aa} & = & U_{\mathbf{q},\mathbf{k}}^{bb}=\frac{U_{3}}{4}\left(-1+\cos{2\theta_{\mathbf{q}}}\cos{2\theta_{\mathbf{k}}}+s\sin{2\theta_{\mathbf{q}}}\sin{2\theta_{\mathbf{k}}}\right)\\
U_{\mathbf{q},\mathbf{k}}^{ab} & = & U_{\mathbf{q},\mathbf{k}}^{ba}=-\frac{U_{3}}{4}\left(1+\cos{2\theta_{\mathbf{q}}}\cos{2\theta_{\mathbf{k}}}+s\sin{2\theta_{\mathbf{q}}}\sin{2\theta_{\mathbf{k}}}\right)\end{eqnarray*}
 and we defined $s\equiv U_{1}/U_{3}$.

We see from (\ref{H_pair}) that both intra-pocket and inter-pocket
pairing interactions acquire angular dependence via the $\cos{2\theta_{\mathbf{q}}},\,\cos{2\theta_{\mathbf{k}}}$
and $\sin{2\theta_{\mathbf{q}}},\,\sin{2\theta_{\mathbf{k}}}$ factors.
Constructing the set of linearized BCS gaps equations for the angle-dependent
gaps $\Delta_{a}(\mathbf{q})$ and $\Delta_{b}(\mathbf{q})$ by standard
means we find at $T=T_{c}$ \begin{eqnarray*}
\Delta_{a}(\mathbf{q}) & = & -\int_{a}\frac{dk_{\parallel}}{4\pi^{2}v_{F}}\, U_{\mathbf{q},\mathbf{k}}^{aa}\Delta_{a}(\mathbf{k})L\\
 &  & -\int_{b}\frac{dk_{\parallel}}{4\pi^{2}v_{F}}\, U_{\mathbf{q,k}}^{ab}\Delta_{b}(\mathbf{k})L\\
\Delta_{b}(\mathbf{q}) & = & -\int_{a}\frac{dk_{\parallel}}{4\pi^{2}v_{F}}\, U_{\mathbf{q},\mathbf{k}}^{ba}\Delta_{a}(\mathbf{k})L\\
 &  & -\int_{b}\frac{dk_{\parallel}}{4\pi^{2}v_{F}}\, U_{\mathbf{q,k}}^{bb}\Delta_{b}(\mathbf{k})L\end{eqnarray*}
 where $L\equiv\log\left(\frac{\Lambda}{T_{c}}\right)$, $\Lambda$
is the upper cutoff of the theory, and the Fermi velocity $v_{F}=v_{F}(k_{\parallel})$
varies along $a-$and $b$ Fermi surfaces.

A generic solution of this gap equation is of the form \begin{eqnarray}
\Delta_{a}(\mathbf{q}) & = & g_{1}+g_{2}\cos2\theta_{\mathbf{q}}+g_{3}\sin2\theta_{\mathbf{q}}\nonumber \\
\Delta_{b}(\mathbf{q}) & = & g_{1}-g_{2}\cos2\theta_{\mathbf{q}}-g_{3}\sin2\theta_{\mathbf{q}}\label{gaps}\end{eqnarray}

Note, however, that the three angular components of $\Delta_{a,b}$
are not orthogonal, since e.g. $\int_{a}dq_{\parallel}\cos2\theta_{\mathbf{q}}$
does not vanish. Substituting these expressions into the gap equation,
we re-express it as a matrix equation for $T_{c}$ and $g_{i}$: \begin{equation}
\left(\begin{array}{c}
g_{1}\\
g_{2}\\
g_{3}\end{array}\right)=\frac{U_{3}L}{2}\left(\begin{array}{ccc}
N & N_{c} & N_{s}\\
-N_{c} & -N_{cc} & -N_{cs}\\
-sN_{s} & -sN_{cs} & -sN_{ss}\end{array}\right)\left(\begin{array}{c}
g_{1}\\
g_{2}\\
g_{3}\end{array}\right)\label{Gap_eqn3}\end{equation}
 where \begin{eqnarray}
N & = & 2\int_{a}\frac{dk_{\parallel}}{4\pi^{2}v_{F}}+2\int_{b}\frac{dk_{\parallel}}{4\pi^{2}v_{F}}\nonumber \\
N_{c} & = & 2\int_{a}\frac{dk_{\parallel}}{4\pi^{2}v_{F}}\cos2\theta_{\mathbf{k}}-2\int_{b}\frac{dk_{\parallel}}{4\pi^{2}v_{F}}\cos2\theta_{\mathbf{k}}\nonumber \\
N_{s} & = & 2\int_{a}\frac{dk_{\parallel}}{4\pi^{2}v_{F}}\sin2\theta_{\mathbf{k}}-2\int_{b}\frac{dk_{\parallel}}{4\pi^{2}v_{F}}\sin2\theta_{\mathbf{k}}\nonumber \\
N_{cc} & = & 2\int_{a}\frac{dk_{\parallel}}{4\pi^{2}v_{F}}\cos^{2}2\theta_{\mathbf{k}}+2\int_{b}\frac{dk_{\parallel}}{4\pi^{2}v_{F}}\cos^{2}2\theta_{\mathbf{k}}\nonumber \\
N_{ss} & = & 2\int_{a}\frac{dk_{\parallel}}{4\pi^{2}v_{F}}\sin^{2}2\theta_{\mathbf{k}}+2\int_{b}\frac{dk_{\parallel}}{4\pi^{2}v_{F}}\sin^{2}2\theta_{\mathbf{k}}\nonumber \\
N_{cs} & = & 2\int_{a}\frac{dk_{\parallel}}{4\pi^{2}v_{F}}\cos2\theta_{\mathbf{k}}\sin2\theta_{\mathbf{k}}\nonumber \\
 &  & +2\int_{b}\frac{dk_{\parallel}}{4\pi^{2}v_{F}}\cos2\theta_{\mathbf{k}}\sin2\theta_{\mathbf{k}}\label{dec_7_4}\end{eqnarray}

\begin{figure}[htp]
\includegraphics[width=0.4\textwidth]{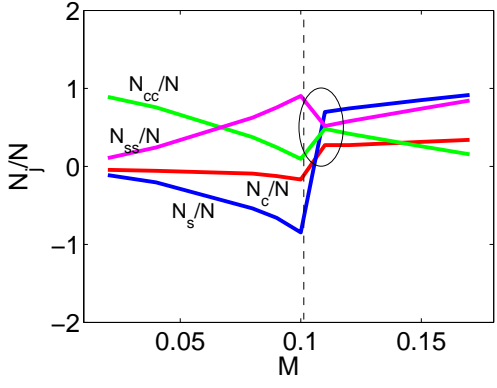}

\caption{\label{fig:N} The quantities $N_{j}/N$, where $N_{j}=N_{c},N_{s},N_{cc}$,
$N_{ss}$, plotted as functions of $M$. The dashed line indicates
the value $M=M_{1}$ at which the $a$-pockets disappear. The ellipsis
indicate the area around $M_{1}$ where the ratios rapidly evolve.
Note that $N_{c}/N$ remains small for all M, while $N_{s}/N$ and
$N_{ss}$ increase with $M$ and become positive and of order one
at $M\geq M_{1}$.}

\end{figure}

\begin{figure}[htp]
 \includegraphics[width=0.23\textwidth]{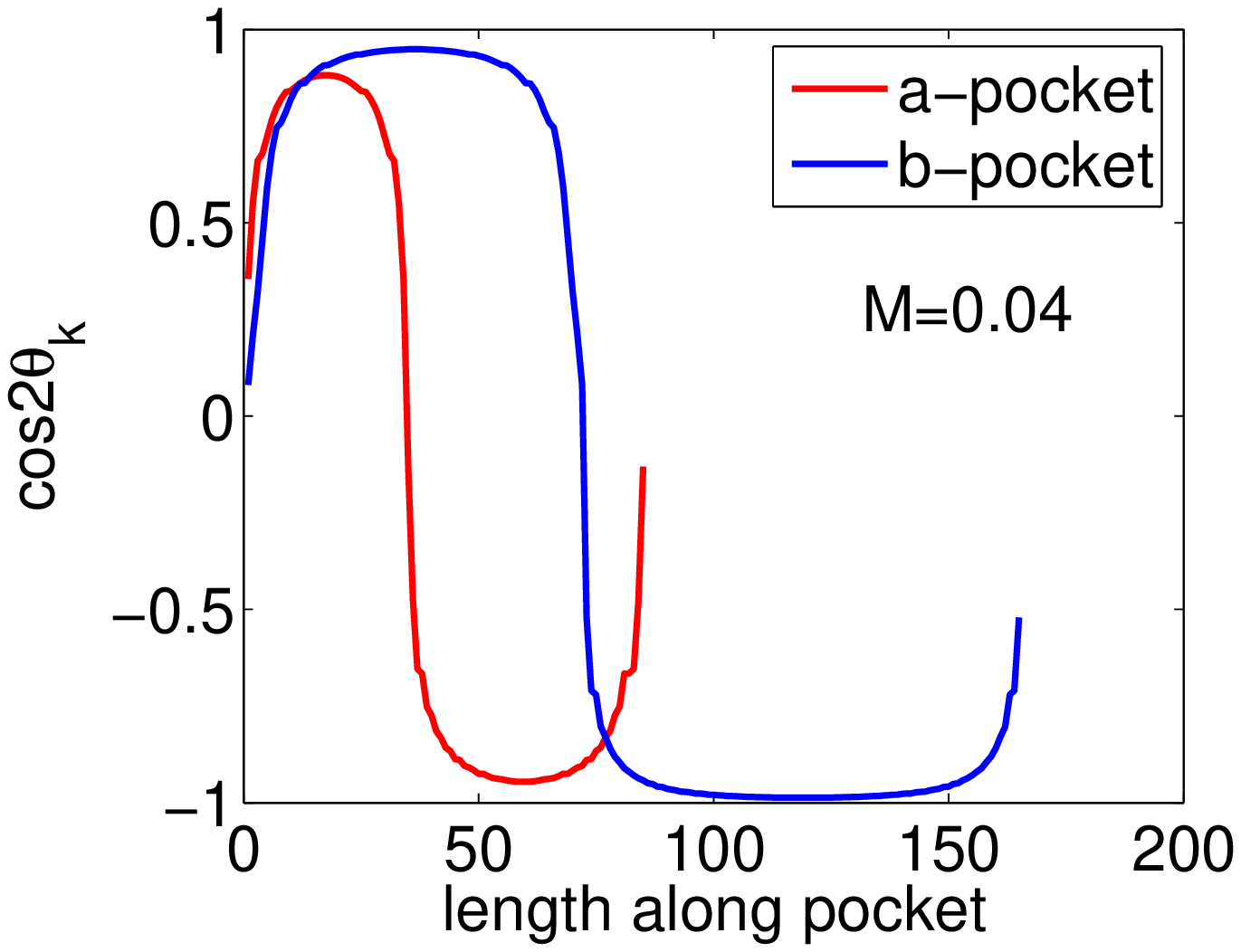} \includegraphics[width=0.23\textwidth]{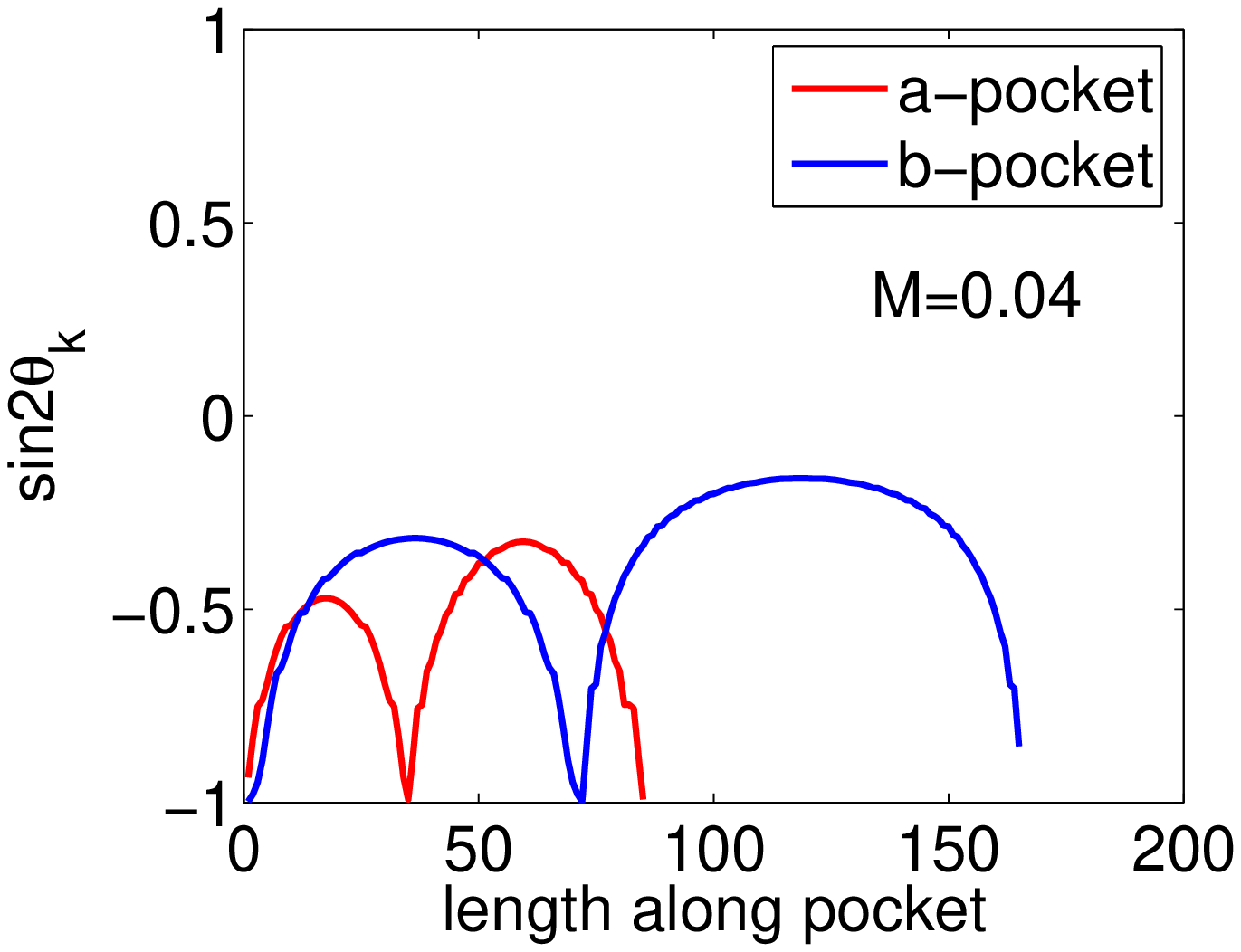}
\includegraphics[width=0.23\textwidth]{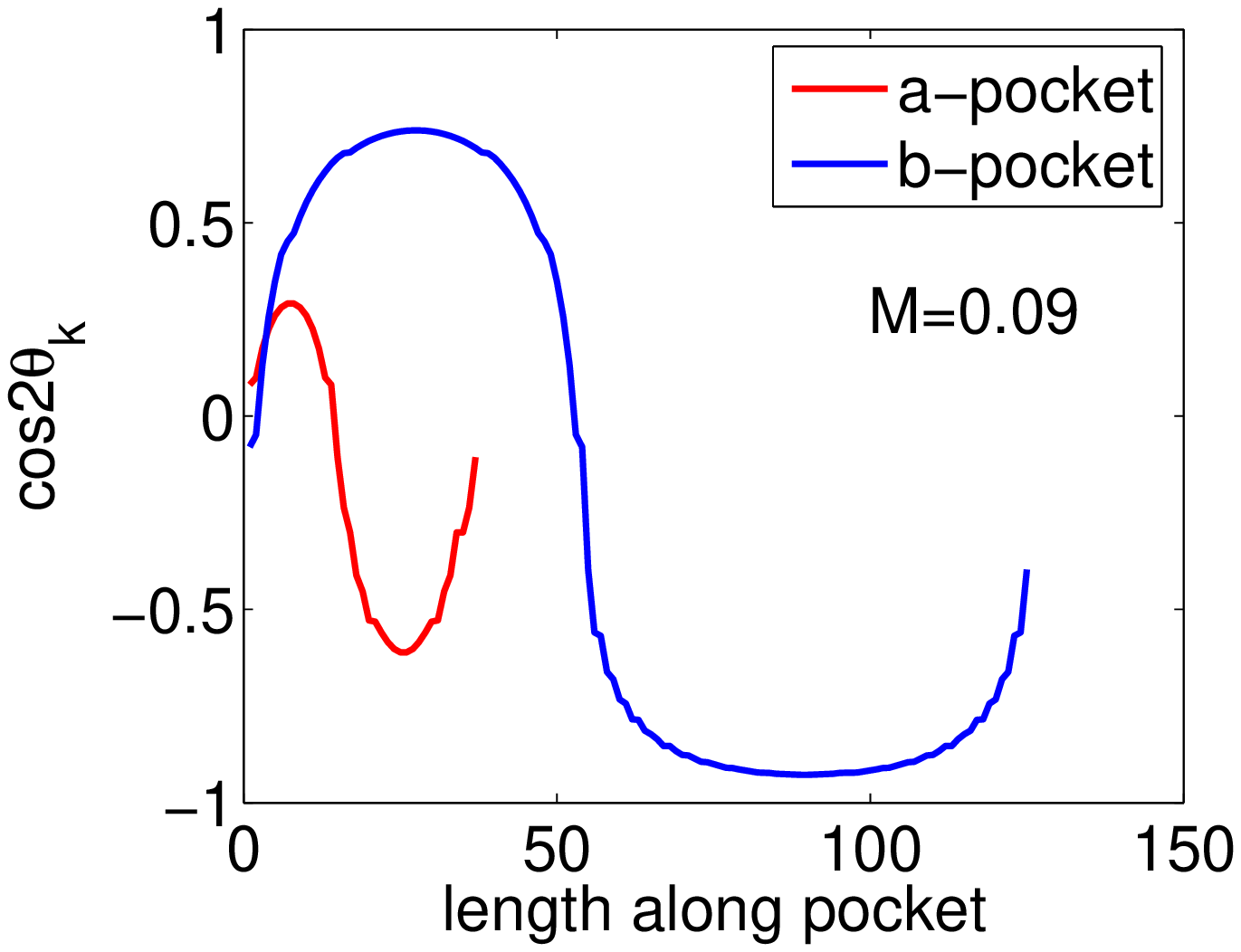} \includegraphics[width=0.23\textwidth]{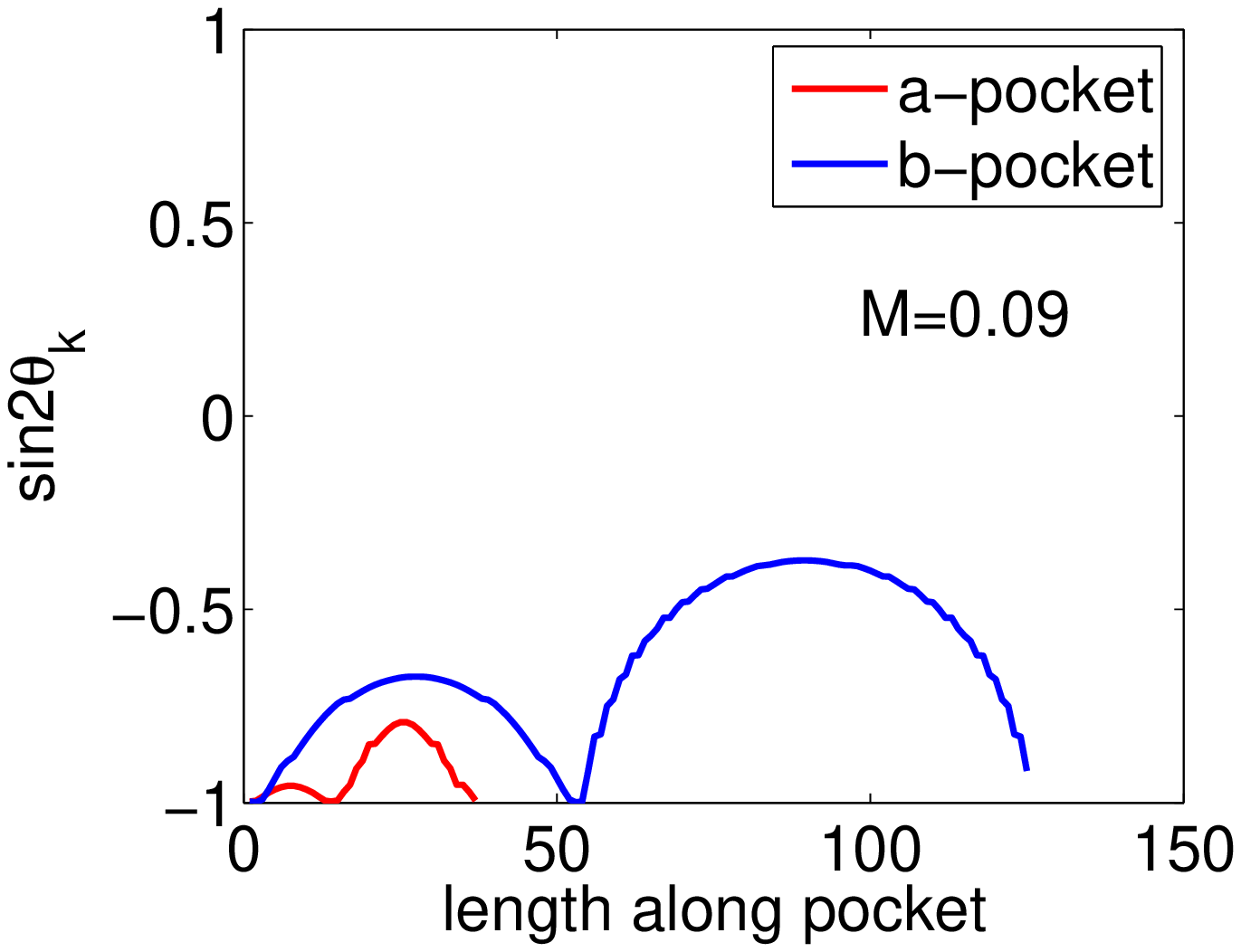}

\caption{\label{fig:cos-sin} The behavior of $\cos{2\theta_{\mathbf{k}}}$
(left column) and $\sin{2\theta_{\mathbf{k}}}$ (right column) along
the $a$ and $b$ pockets for $M=0.04$ and $M=0.09$. The gap is
given by Eq. (\ref{gaps}). The \textquotedbl{}length along the
pocket\textquotedbl{} refers to the numbers along each pocket specified
in Fig. \ref{fig:FS}.}

\end{figure}

Without SDW order, $\sin2\theta_{k}=0$, $\cos2\theta_{k}=\mathrm{sign}(\epsilon_{c}-\epsilon_{f})$,
and $N_{s}$, $N_{sc}$, and $N_{ss}$ all vanish. Then $g_{3}=0$,
while $g_{2}/g_{1}=-N_{c}/(N+\sqrt{N^{2}-N_{c}^{2}})$. As a result,
the gaps on the inner and outer sides of the $a$ and $b$ pockets
reduce to $g_{1}(1+N_{c}/(N+\sqrt{N^{2}-N_{c}^{2}}))$ and $g_{1}(1-N_{c}/(N+\sqrt{N^{2}-N_{c}^{2}}))$.
These two coincide with the gaps on the $c$ and $f-$Fermi surfaces
at $T_{c}$ in the paramagnetic phase, which we set in our model to
be angle-independent (we recall that the sign of the two gaps is the
same because we attributed an extra minus sign to the gap on the $f-$Fermi
surface by flipping the spins of $f-$fermions). The magnitudes of
the gaps on $c$ and $f-$ FSs are generally unequal as $N_{c}\neq0$
because of the different geometry of hole and electron pockets. Numerically,
however, for our set of parameters $N_{c}/N\ll1$, i.e., at $M=0$
the gaps on $a$ and $b$ FSs are essentially isotropic. Once SDW
order sets in, all $N_{i}$ and $N_{ij}$ become non-zero, $g_{3}$
becomes finite, $\cos2\theta_{\mathbf{k}}$ and $\sin2\theta_{\mathbf{k}}$
become smooth functions along the pockets, and the gaps $\Delta_{a}({\bf \mathbf{q}})$
and $\Delta_{b}(\mathbf{q})$ acquire smooth angular variations. The
magnitudes of the variations depend on the two parameters: the ratios
$g_{2}/g_{1}$ and $g_{3}/g_{1}$ and the actual variation of $\cos2\theta_{\mathbf{k}}$
and $\sin2\theta_{\mathbf{k}}$ along $a-$and $b-$ Fermi surfaces.

In the small $M$ limit, considered previously in Ref.
\onlinecite{parker_09}, $\sin{2\theta_{\mathbf{k}}}$ is small
except for narrow ranges near the tips of the bananas. Then
$N_{s}$, $N_{sc}$, and $N_{ss}$ are small, and $N_{cc}/N\approx1$.
Solving Eq. (\ref{Gap_eqn3}) in this limit, we find that $g_{3}\ll
g_{2}\ll g_{1}$, hence $\Delta_{a}$ and $\Delta_{b}$ remain almost
constant along the $a-$ and $b-$pockets, including the tips of the
bananas.

\begin{figure}[htp]
 \includegraphics[width=0.45\textwidth]{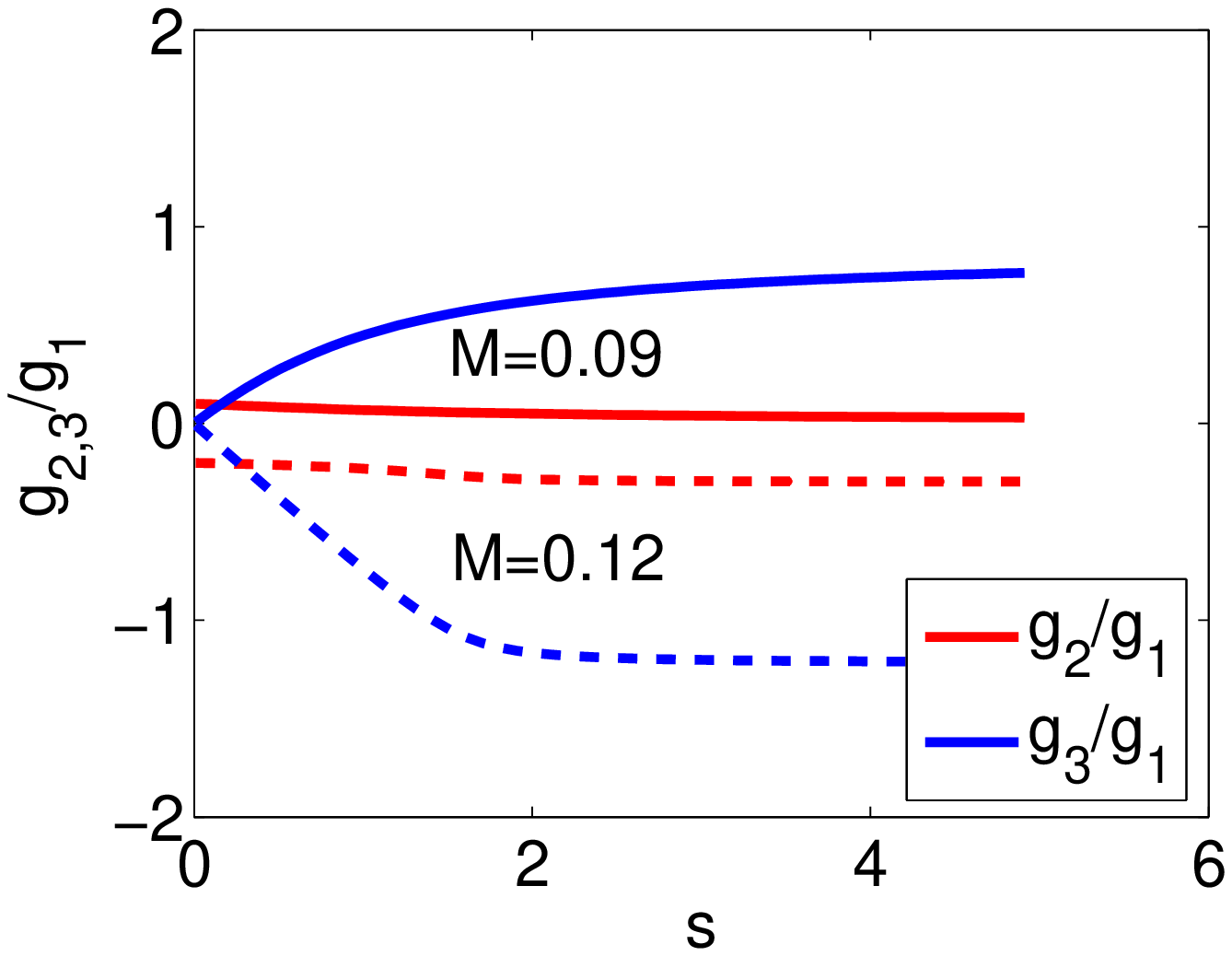}

\includegraphics[width=0.45\textwidth]{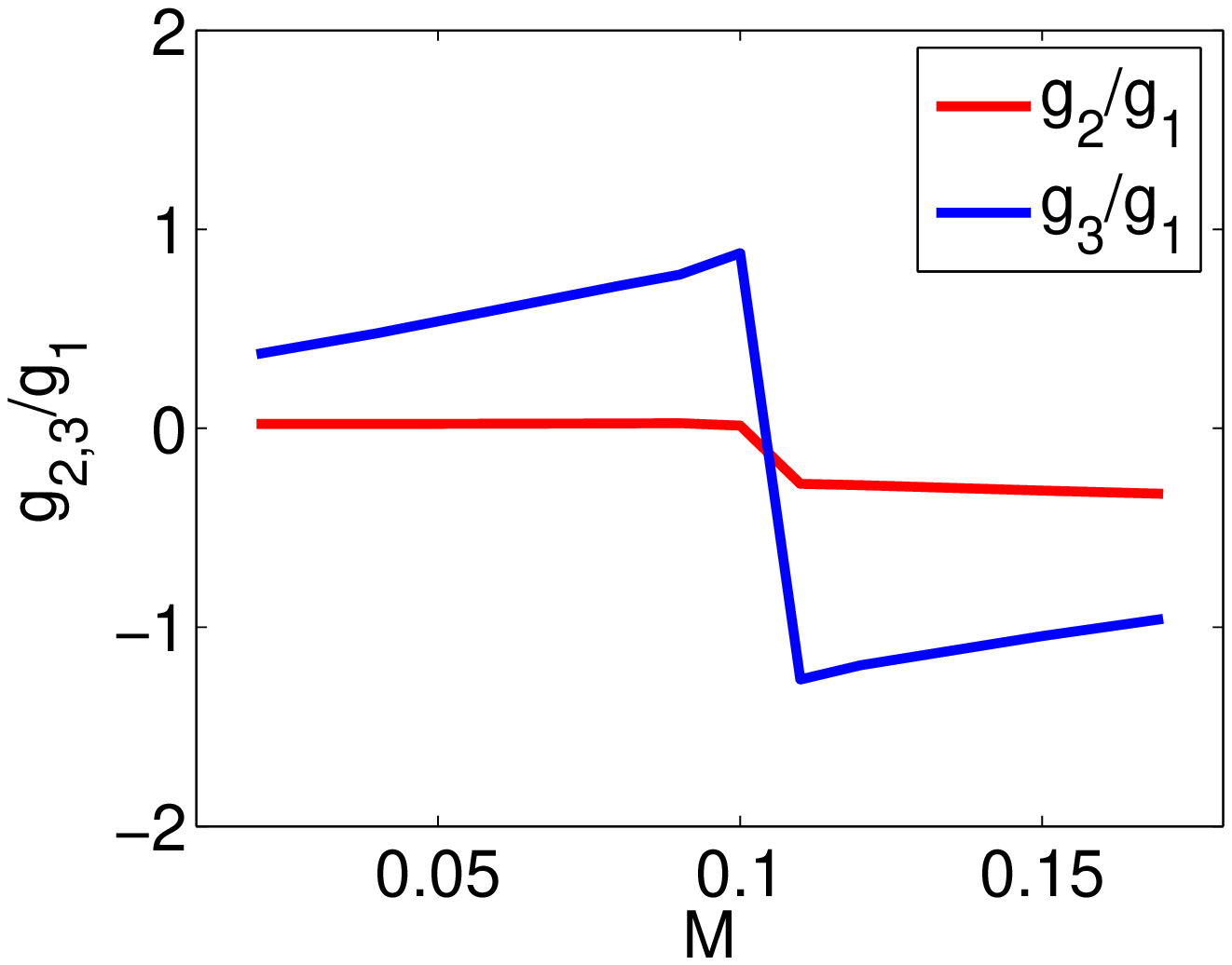}

\caption{\label{fig:g_plots} Top panel: the ratios $g_{2}/g_{1}$ (blue line)
and $g_{3}/g_{1}$ (red line) as function of $s$ for $M=0.09$ (solid
lines) and $M=0.12$ (dashed lines). These values are slightly below
and slightly above the critical $M_{1}=0.102$ at which two out of
four pockets disappear. Both ratios flip sign between $M=0.09$ and
$M=0.12$. The key result here is that for $s\geq1$, the magnitude
of $g_{3}/g_{1}$ becomes larger than $1$ for $M=0.12$. This gives
rise to accidental nodes of the gap on the remaining FS pocket (see
text). Bottom panel: the plot of the same ratios as functions of $M$
for $s=5$. The magnitude of $g_{2}/g_{1}$ remains relatively small,
while the magnitude of $g_{3}/g_{1}$ becomes greater than 1 for $M\geq M_{1}$.}

\end{figure}

For arbitrary $M$, $g_{2}/g_{1}$ and $g_{3}/g_{1}$ are rather complex
functions of the ratios $N_{j}/N$ and also $s$. We plot $N_{j}/N$
in Fig \ref{fig:N}. We see that $N_{c}/N$ and $N_{cs}/N$, are small
because the integrands for $N_{c}$ and $N_{cs}$ contain $\cos2\theta_{k}$,
which changes sign between the inner and outer sides of the $a$ and
$b$-pockets (Fig. \ref{fig:cos-sin}). At the same time, the magnitudes
of $N_{s}/N$ and $N_{ss}/N$ increase with increasing $M$ and become
of order one.

In Fig. \ref{fig:cos-sin}, we plot $\cos{2\theta_{\mathbf{k}}}$
and $\sin{2\theta_{\mathbf{k}}}$ along the $a$ and $b$-pockets.
We recall that the prefactors of the $\cos2\theta_{\mathbf{k}}$ terms
in $\Delta_{a}$ and $\Delta_{b}$ are already small because $g_{2}/g_{1}$
is small. Fig. \ref{fig:cos-sin} shows that the variation of $\cos{2\theta_{\mathbf{k}}}$
along each of the pockets decreases with increasing $M$, making the
$g_{2}\cos2\theta_{\mathbf{k}}$ term even less relevant. On the other
hand, the range where $|\sin{2\theta_{\mathbf{k}}}|$ is not small
widens up with increasing $M$. Over some range of $M$, up to $M_{11}=0.5(m_{y}-m)/(m_{y}+m)<M_{1}$
on the $a$-pocket and up to $M_{22}=0.5(m-m_{x})/(m+m_{x})<M_{2}$
on the $b$-pocket, $\sin{2\theta_{\mathbf{k}}}$ varies between $-1$
and some negative value. At $M_{11}<M<M_{1}$, $\sin{2\theta_{\mathbf{k}}}$
on the $a$-pocket varies in a more narrow interval, does not reach
$-1$ and eventually becomes

\begin{equation}
\sin{2\theta_{\mathbf{k}_{a}}}_{M\to M_{1}}=-2\frac{\sqrt{mm_{y}}}{m+m_{y}}=-0.98.\end{equation}

The same behavior holds for $\sin{2\theta_{\mathbf{k}}}$ on the $b-$pocket
for $M_{22}<M<M_{2}$, but with

\begin{equation}
\sin{2\theta_{\mathbf{k}_{b}}}_{M\to M_{2}}=-2\frac{\sqrt{mm_{x}}}{m+m_{x}}=-0.94.\label{sin_theta_b}\end{equation}

\section{Gap structure in the coexistence state as function of $M$}

\label{sec:results}

\subsection{From small to intermediate $M$: nodeless-nodal transition}

We now discuss in more detail the solution of the matrix equation
(\ref{Gap_eqn3}) and the structure of the SC gap. We found in the
previous section that the angular dependence of $\cos{2\theta_{\mathbf{k}}}$
and $\sin{2\theta_{\mathbf{k}}}$ along the $a$ and $b-$ pockets
depends on $M$, while the solution of the matrix equation for $g_{i}$
depends on $M$ and also on $s$ - which, we remind, is the ratio
of the interactions $s=U_{1}/U_{3}$.

The solution of the matrix equation can be easily analyzed in the
limits of small and large $s$. At small $s$, $g_{3}$ is small compared
to $g_{2}$ ($g_{3}/g_{2}=O(s)$), while $g_{2}/g_{1}$ is given by

\begin{equation}
\frac{g_{2}}{g_{1}}\approx-\frac{N_{c}}{N+N_{cc}}\label{g2/g1}\end{equation}

This ratio is always small because $N_{c}/N$ is small for all $M$
(see Fig. \ref{fig:N}). As a result, the gaps $\Delta_{a}(q)$ and
$\Delta_{b}(q)$ remain essentially constants regardless of how strong
the SDW order is.

In the opposite limit $s\gg1$ the behavior is different. Only in
a narrow range of the smallest $M\ll1/s$ the gaps remain almost angle-independent.
Outside this range $g_{2}$ is small compared to $g_{3}$ ($g_{2}/g_{3}=O(1/s)$),
while $g_{3}/g_{1}$ is given by

\begin{equation}
\frac{g_{3}}{g_{1}}\approx-\frac{N_{s}}{N_{ss}}\label{g3/g1}\end{equation}

Both $N_{s}$ and $N_{ss}$ become non-zero at a finite $M$ and their
ratio is of order one, i.e., $g_{3}\sim g_{1}$. This leads to sizable
angular dependencies of $\Delta_{a}(q)$ and $\Delta_{b}(q)$. For
small $M<M_{1}$, $N_{s}$ is smaller in magnitude than $N_{ss}$
(see Fig. \ref{fig:N}), hence $|\frac{g_{3}}{g_{1}}|<1$ and the
angular variations of the gaps do not give rise to nodes. However,
as $M$ increases and approaches $M_{1}$, the ratio $N_{s}/N_{ss}$
changes sign, becomes negative and its magnitude exceeds one (see
Fig. \ref{fig:N}). The combination of this behavior and the fact
that for $M\sim M_{1}$ $\sin2\theta$ reaches $-1$ at the tip of
the bananas implies that the gap along the $b-$ pocket develops accidental
nodes. This happens even before $M$ reaches $M_{1}$ because at $M=M_{1}$
(when the $a-$pocket disappears), $|N_{s}/N_{ss}|$ is already larger
than one, as one can immediately see from Eq. \ref{dec_7_4}, if one
neglects in this equation the contributions from the $a$ pocket.
The nodes in $\Delta_{b}(\mathbf{q})$ obviously survive up to $M_{22}$
because at $M<M_{22}$ the minimal value of $\sin{2\theta_{\mathbf{q}}}$
remains $-1$. At larger $M_{22}<M<M_{2}$ this minimal value becomes
smaller than one, but we checked numerically that the nodes still
survive and remain present up to $M=M_{2}$.

To understand the gap structure in between the limits $s\ll1$ and
$s\gg1$, we solve the $3\times3$ gap equation (\ref{Gap_eqn3})
numerically for several $s$ with $M$ as a running parameter, and
for several $M$ with $s$ as a running parameter. In the two panels
in Fig. \ref{fig:g_plots} we plot the ratios $g_{2}/g_{1}$ and $g_{3}/g_{1}$
as function of $s$ for two representative values of $M$, $M=0.09<M_{1}$
and $M_{1}<M=0.12<M_{2}$, and as function of $M$ for $s=5$. We
see that $\left|g_{2}/g_{1}\right|$ is always small, while $\left|g_{3}/g_{1}\right|$
increases with increasing $s$ and for $M=0.12$ becomes larger than
$1$ above a certain $s$. Once this happens, the gap on the $b-$pocket
develops accidental nodes. Overall, these results show that the behavior
that we obtained analytically at large $s$ extends to all $s\geq1$.
For these values of $s$, the gap along the $b-$pocket, which survives
up to larger $M$, necessarily develops accidental nodes around $M=M_{1}$,
where the $a$-pocket disappears.

\begin{figure*}[htp]
 \includegraphics[width=0.3\textwidth]{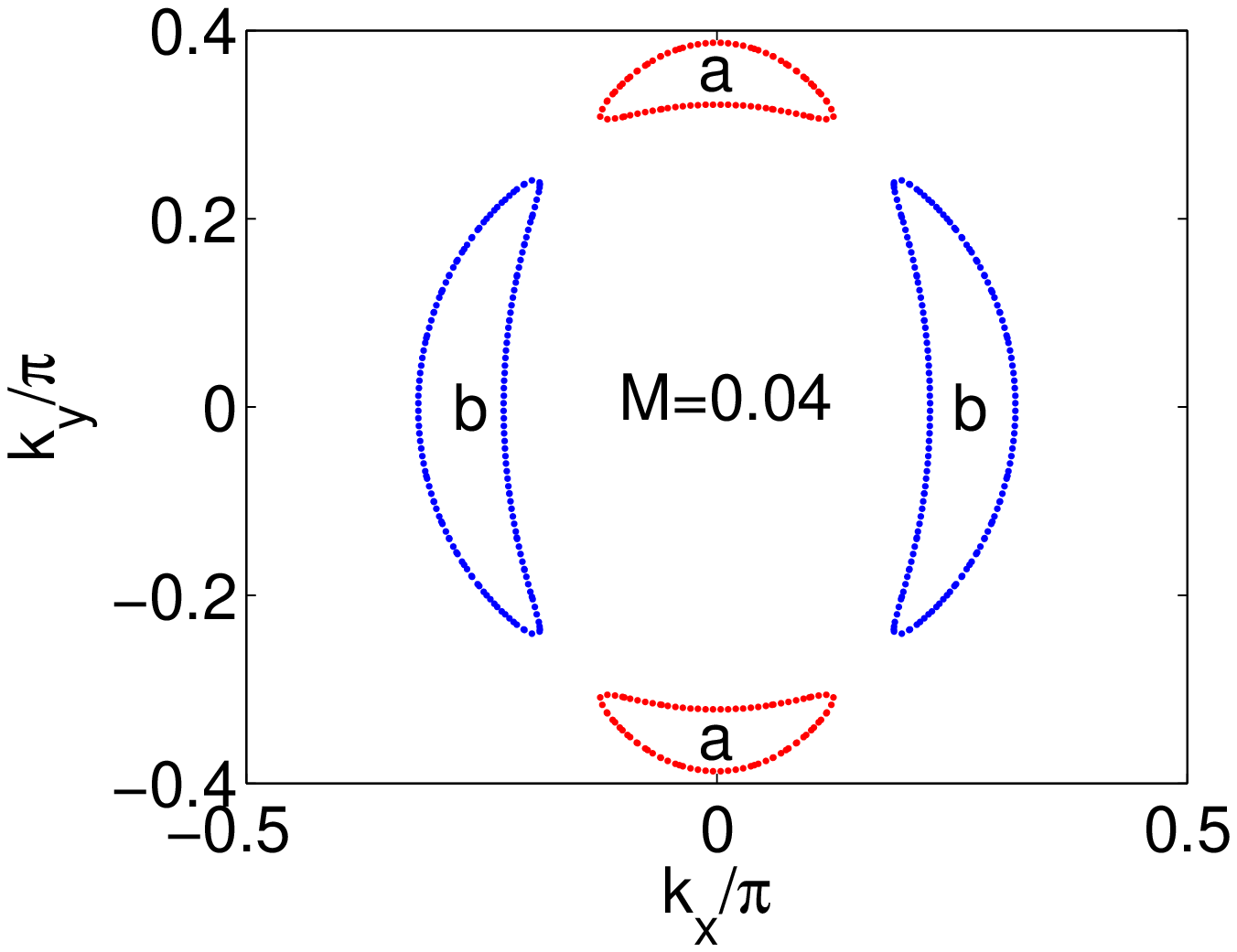} \includegraphics[width=0.3\textwidth]{FS_Dp09_2poc}
\includegraphics[width=0.3\textwidth]{FS_Dp12_2poc}\\
 \includegraphics[width=0.3\textwidth]{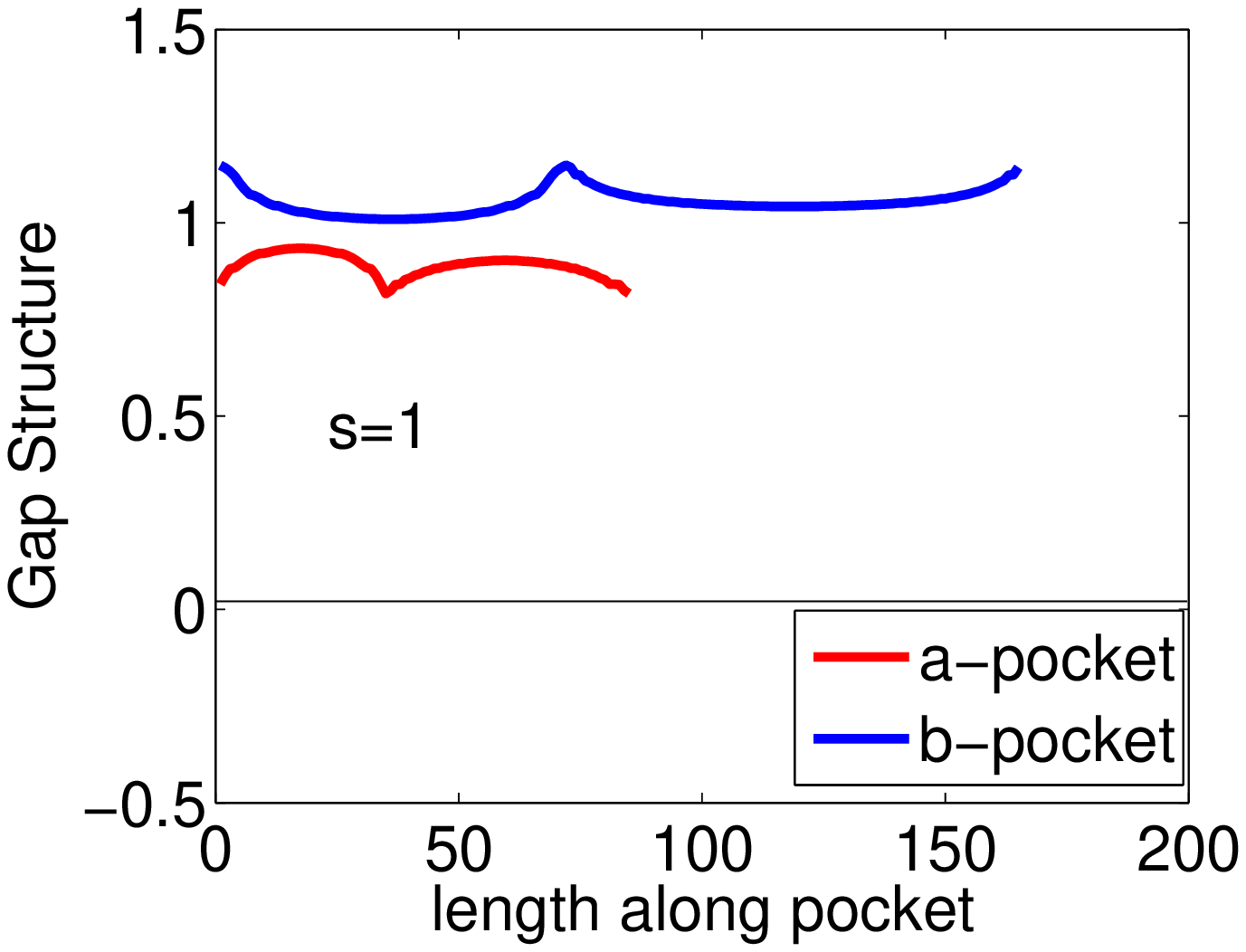} \includegraphics[width=0.3\textwidth]{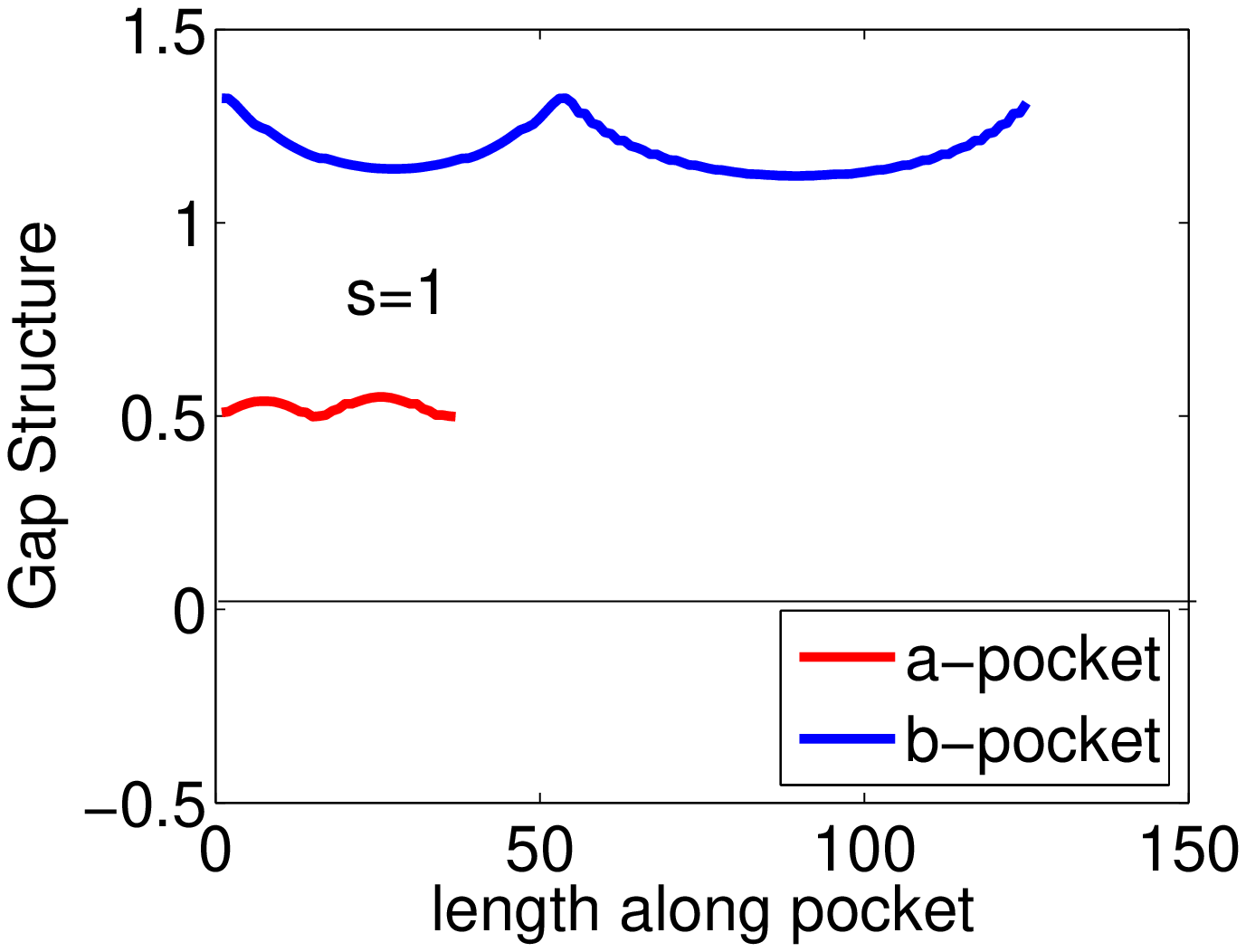}\&
\includegraphics[width=0.3\textwidth]{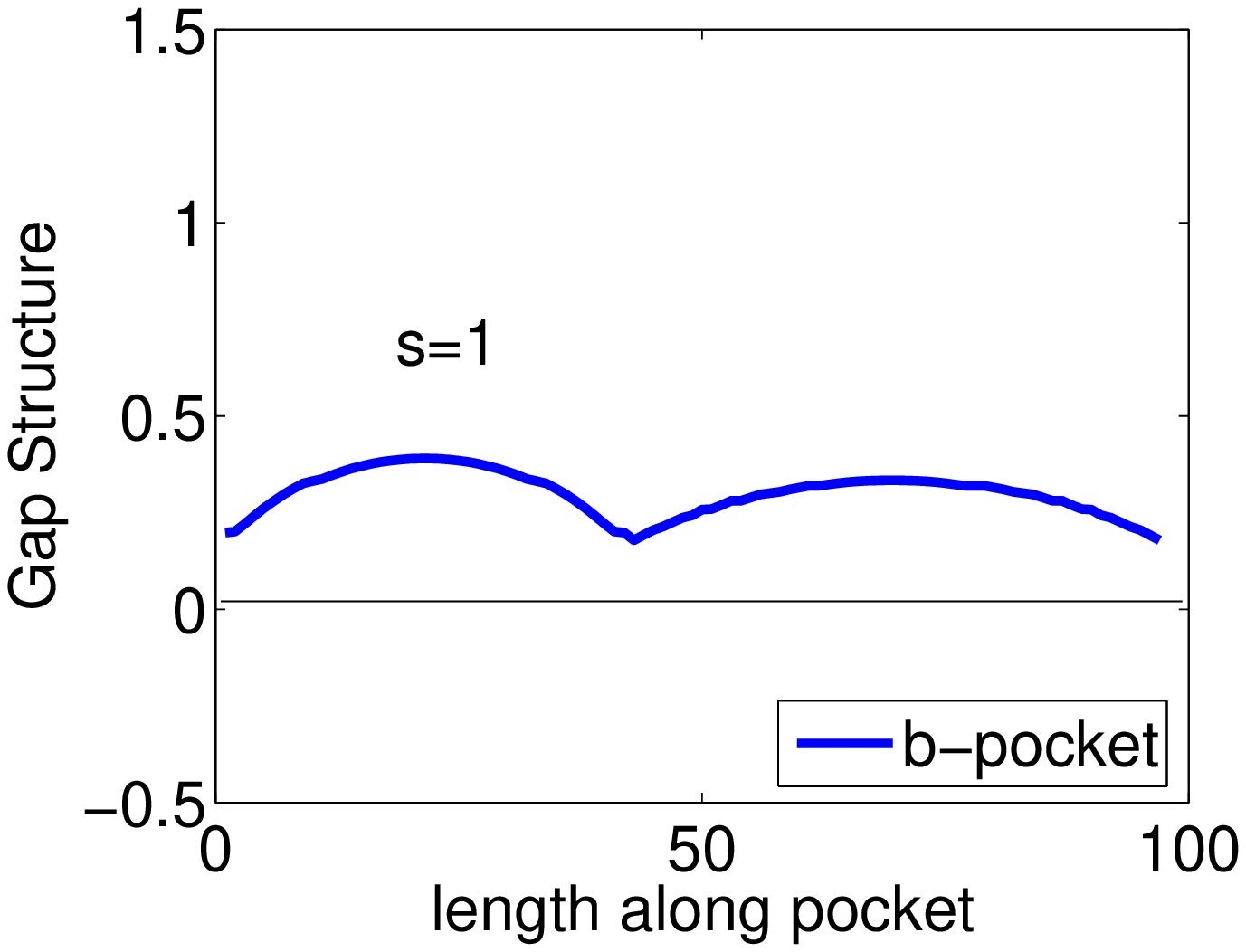}\\
 \includegraphics[width=0.3\textwidth]{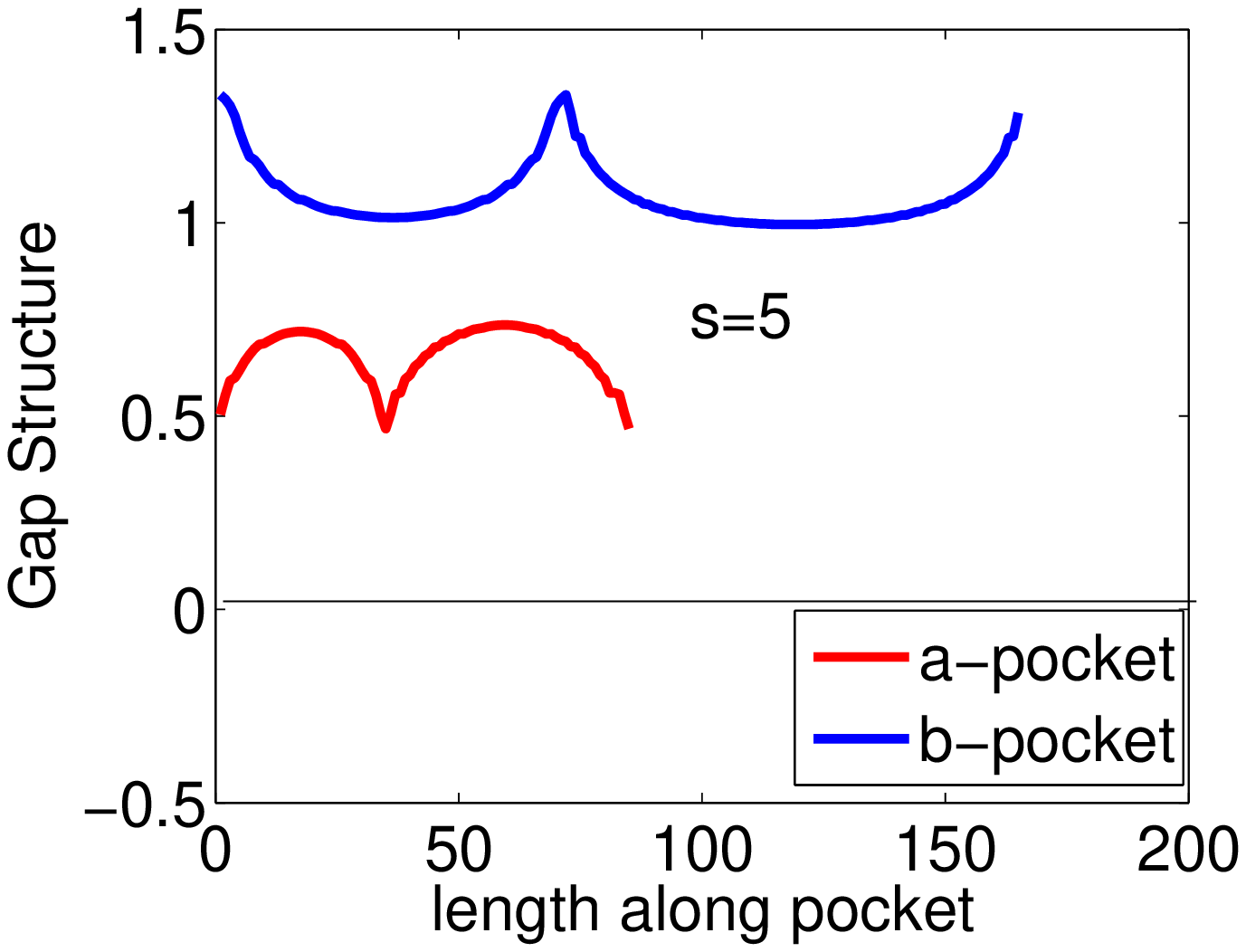} \includegraphics[width=0.3\textwidth]{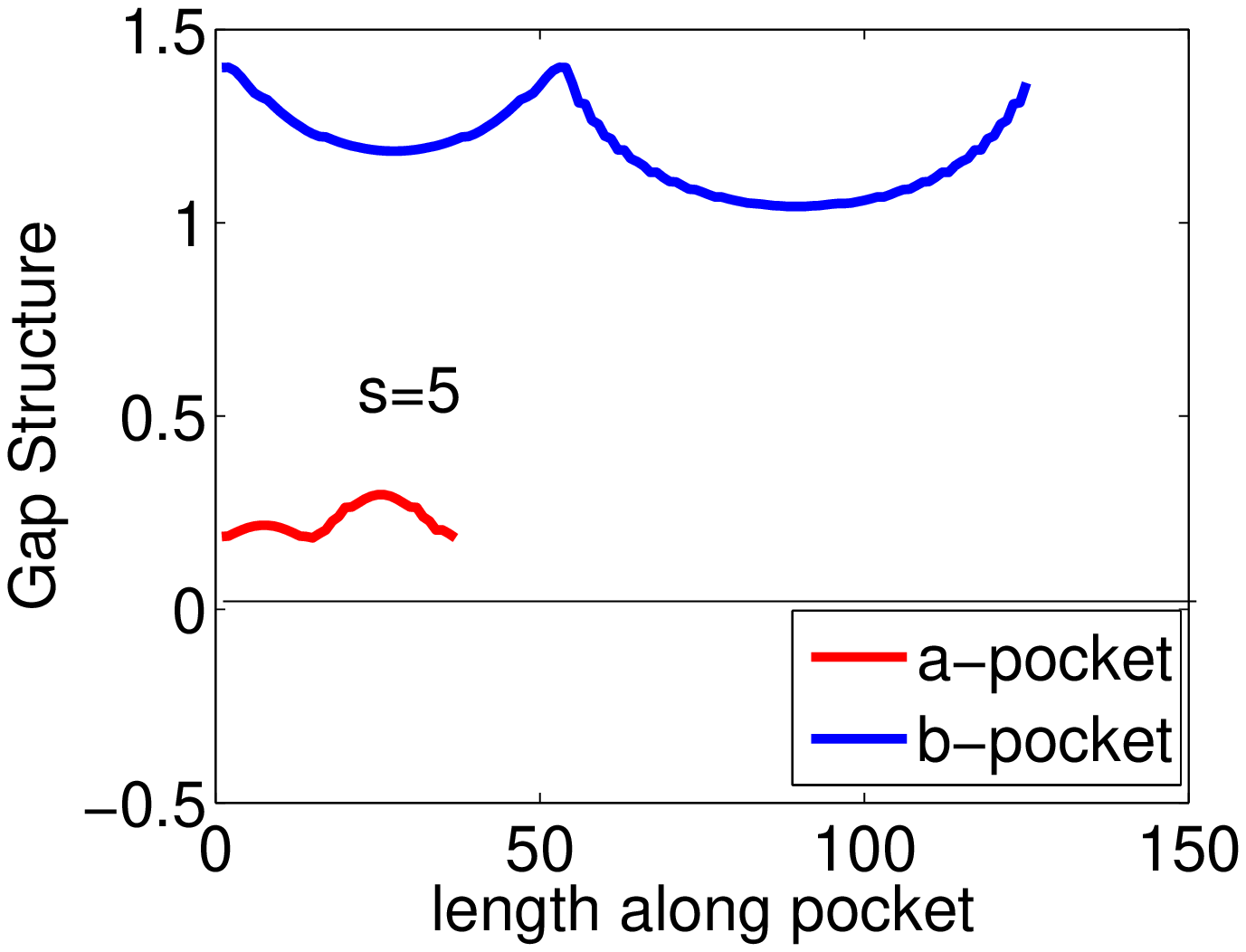}
\includegraphics[width=0.3\textwidth]{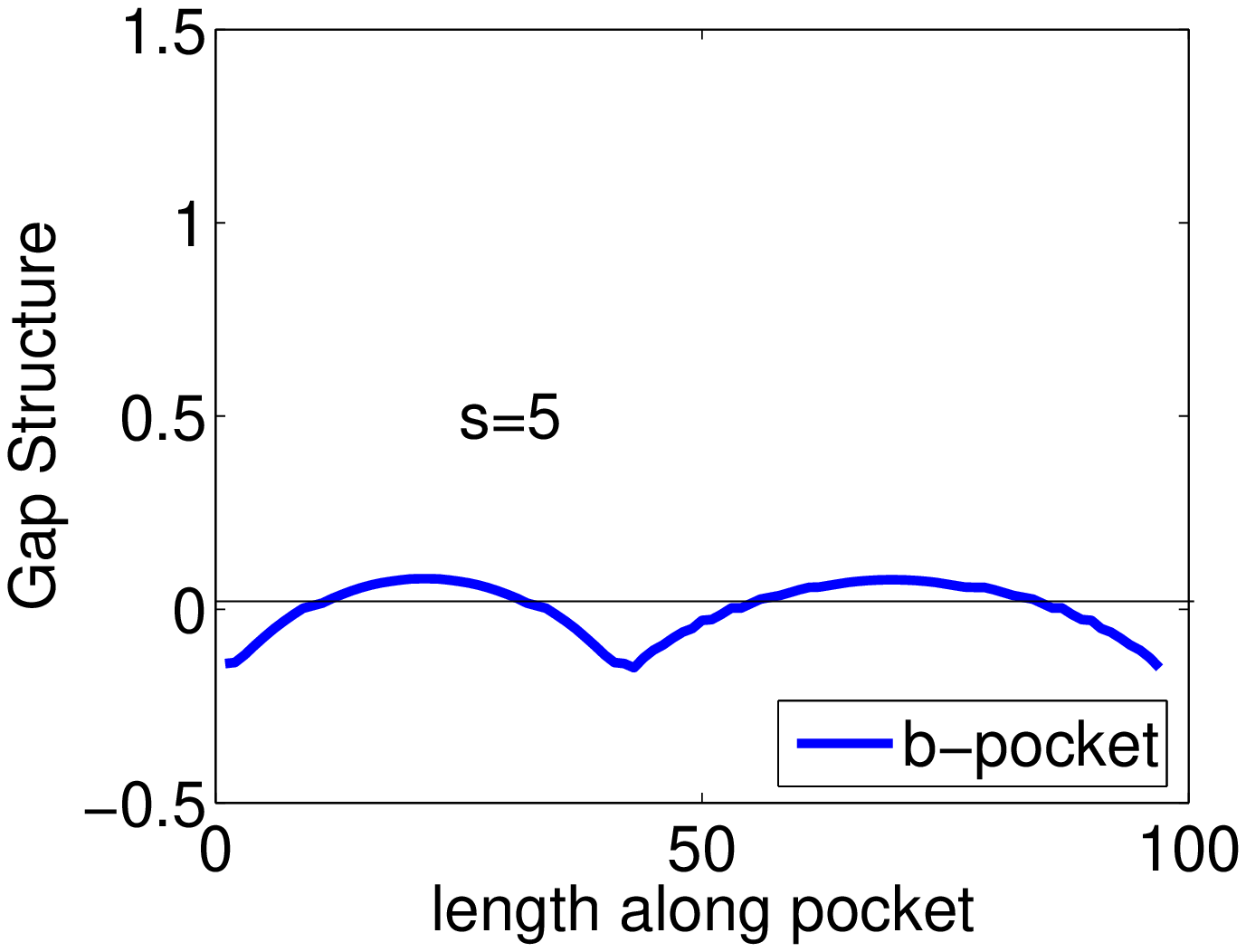}

\caption{\label{fig:FS_gaps} The gap structure for two values of $s=U_{1}/U_{3}$
and three different $M$ corresponding to the cases when the $a$
and $b$-pockets are of comparable size (left column), when the $a$
pockets are about to disappear (middle column), and when only the
$b-$pockets are left (right column). Note the appearance of gap nodes
for $s=5$ and $M=0.12$, when one pair of reconstructed pockets disappear.}

\end{figure*}

In Fig. \ref{fig:FS_gaps} we show the variations of $\Delta_{a}(\mathbf{q})$
and $\Delta_{b}(\mathbf{q})$ for two different values of $s$, $s=1$
and $s=5$, and three different values of $M$: $M\ll M_{1}$, $M\leq M_{1}$,
and $M_{1}<M<M_{2}$. For $s=1$, the gaps $\Delta_{a}$ and $\Delta_{b}$
have no nodes for all $M$, consistent with $|g_{3}/g_{1}|<1$ in
Fig. \ref{fig:g_plots}. The only effect of the disappearance of the
$a-$pockets is the switch between the maxima and the minima of the
gap function along the remaining $b$-pockets. This switch is a consequence
of the sign change of both $g_{2}/g_{1}$ and $g_{3}/g_{1}$ (see
Fig. \ref{fig:g_plots}).

The situation is different for larger $s=5$. Even for small $M\ll M_{1}$,
the gap variation along both $a-$and $b-$pockets become substantial.
Once $M$ becomes large enough to (almost) gap out the $a$-pockets,
the nodes appear near the tips of the $b-$pocket bananas. This behavior
is in full agreement with the analysis of $g_{3}/g_{1}$ earlier in
this section.

We therefore conclude that the two conditions to obtain nodes in
the superconducting gap in the coexistence phase with stripe
antiferromagnetism are: (i) relatively large density-density
interaction $U_{1}$ (leading to magnetism) compared to the
pair-hopping interaction $U_{3}$, and (ii) the disappearance of
one of the two pairs of reconstructed Fermi pockets. Our analysis
agrees with previous works that found fully gapped quasi-particle
excitations in the coexistence state in the cases of small $M$
(Ref. \onlinecite{parker_09}) and $U_{1}=0$ (Ref.
\onlinecite{rafael}).

\begin{figure*}[htp]
 \includegraphics[width=0.3\textwidth]{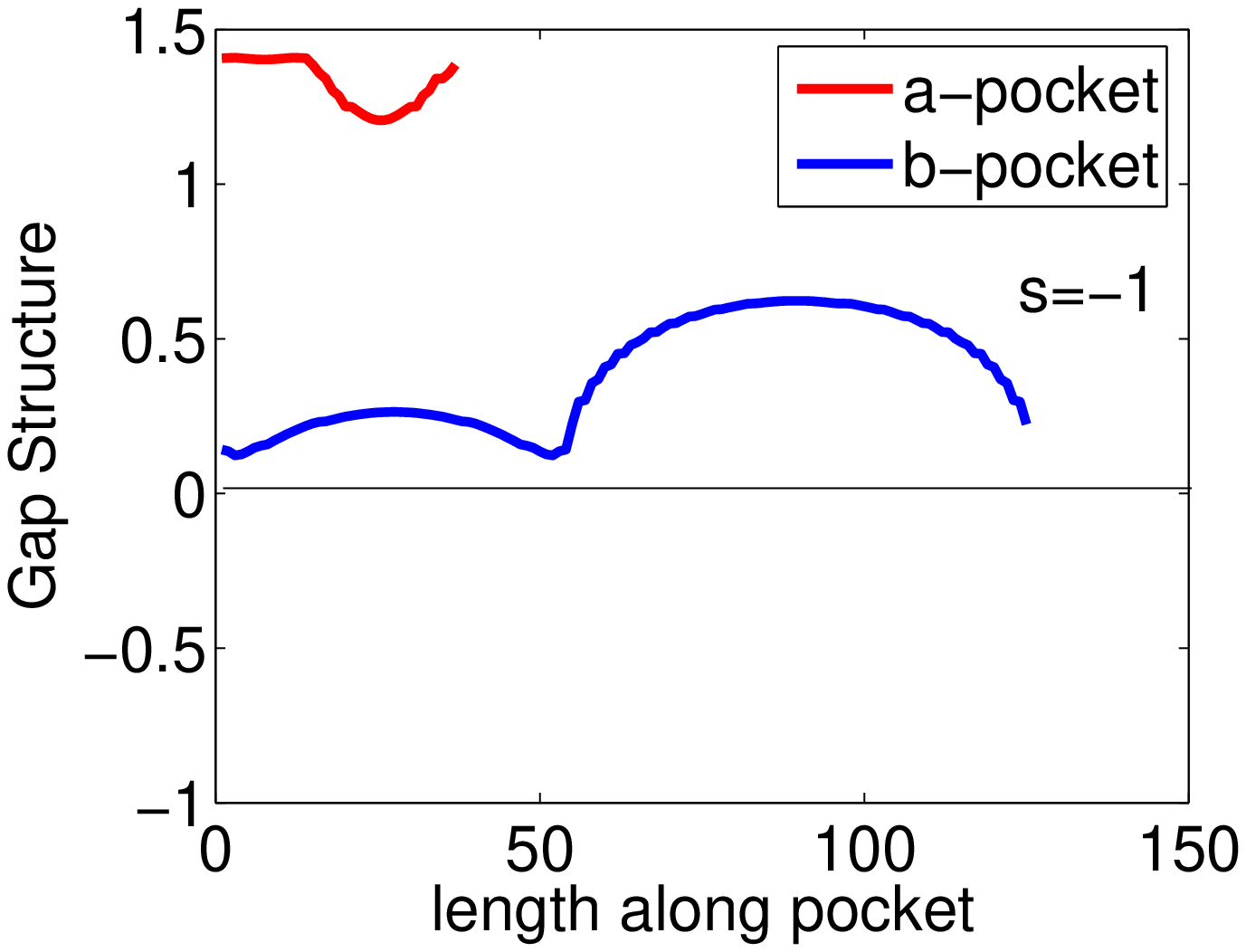} \includegraphics[width=0.3\textwidth]{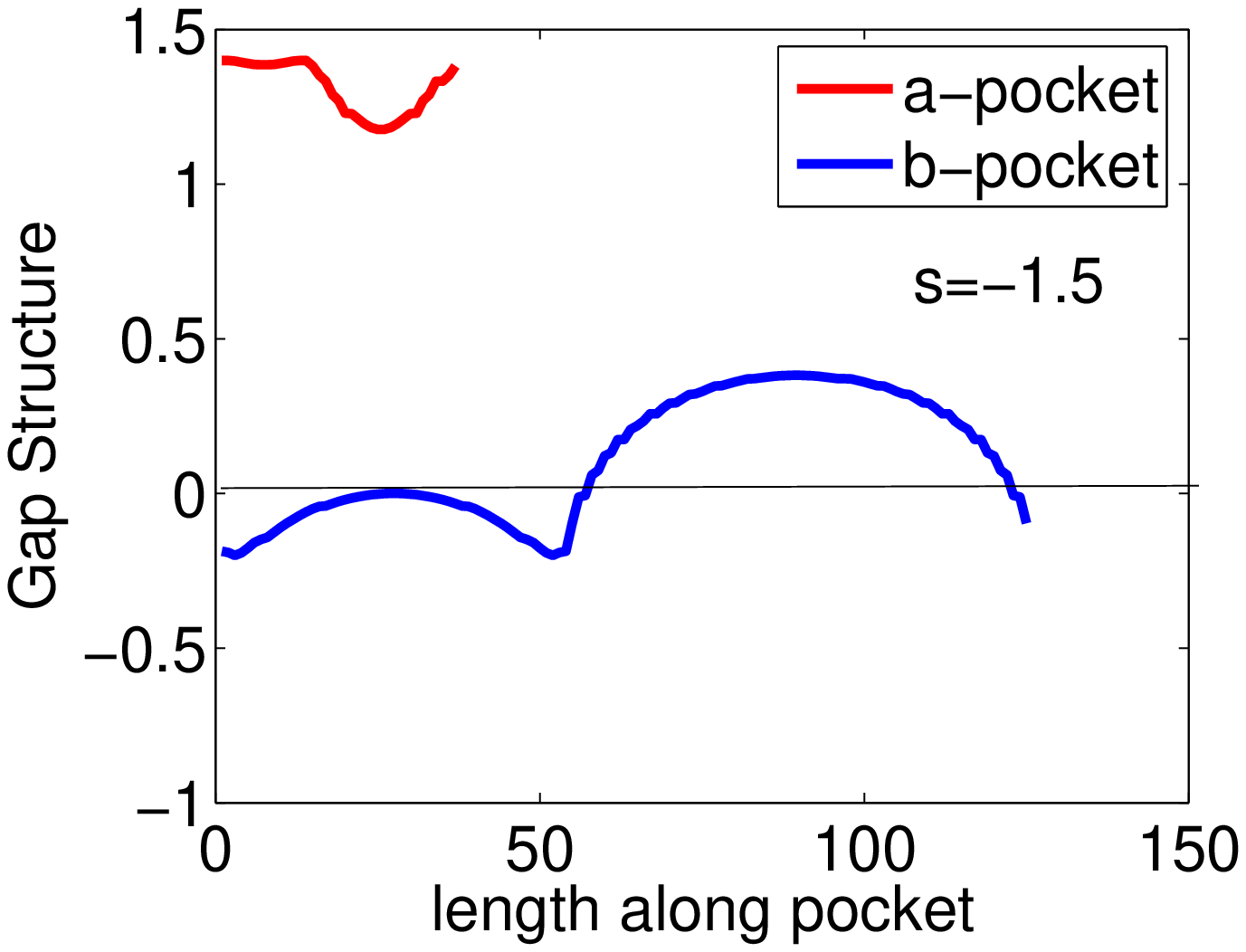}
\includegraphics[width=0.3\textwidth]{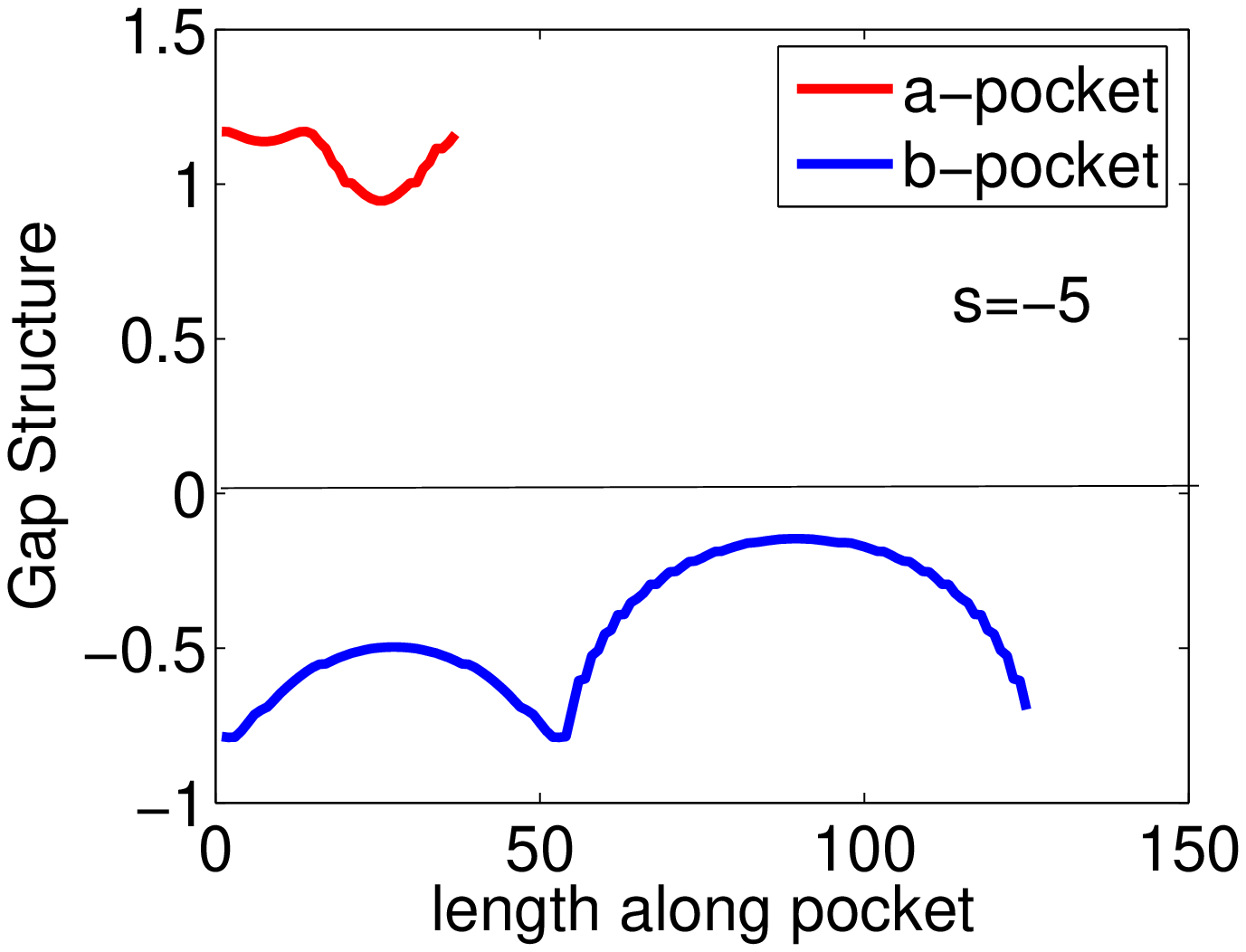}

\caption{\label{fig:nn_transition} Gap structure for negative $s$ and $M=0.09$.
As $|s|$ increases, the gap on the $b$ pocket is pushed down, acquires
the nodes, and eventually reverses sign and again becomes nodeless.
For all three values of $s$ the FS contains both $a$ and $b$ pockets.}

\end{figure*}

The other interactions, which we listed in Sec. II but did not include
into $H_{int}$ in Eq. (\ref{1}), modify the value of $s$ and therefore
affect whether or not the nodes appear upon increasing $M$. In particular,
the exchange term $\left[U_{2}c_{\sigma}^{\dagger}f_{\sigma}f_{\sigma^{\prime}}^{\dagger}c_{\sigma^{\prime}}\right]$
changes $s$ to $s=\frac{U_{1}-U_{2}}{U_{3}}$. If $U_{2}$ is negative,
it makes the development of nodes more likely, while if it is positive
and smaller than $U_{1}$, it decreases $s$ and may eliminate the
nodes. The inclusion of this term also opens up the somewhat exotic
possibility of a negative $s$. Although negative $s$ is unlikely
for FeSCs, we analyzed the $s<0$ case for completeness and show the
results in Fig. \ref{fig:nn_transition}. We see that now the behavior
is non-monotonic with $s$: the nodes appear at some intermediate
$|s|$ and disappear at larger $|s|$, already at relatively small
$M$, when all four pockets are present.

\subsection{From intermediate to large $M$: nodal-nodeless transition}

In the previous section we showed that nodes appear at large enough
$s$ at intermediate values of $M\sim M_{1}$, where one of the two
pairs of reconstructed pockets (the $a$-pocket) disappears. We now
analyze what happens when $M$ becomes larger than $M_{2}$ and $b$-pockets
also disappear.

We find that the results depend on whether we restrict the pairing
to the FS or include into the pairing problem also the states
which are already gapped out by SDW. If we restrict the pairing
problem to the FS, as we did before, we find that the nodes in
$\Delta_{b}$ are present for all $M<M_{2}$. At $M=M_{2}$ the
remaining FS disappears and $T_{c}$ vanishes. If we do not
restrict the pairing to the FS and take into consideration the
states already gapped by SDW, SC persists into the region
$M>M_{2}$ (see Refs. \onlinecite{vvc_10,maxim,fernandes_optics}).
In this last region, all states are gapped already above $T_{c}$,
and the opening of an additional SC gap only moves states further
away from zero energy. The thermal conductivity and the
penetration depth in this last regime show exponential behavior,
typical for a superconductor with a full gap, From this
perspective, the evolution of the system response around $M=M_{2}$
mimics the transformation from a nodal to a nodeless gap,
irrespective of whether the actual SC gap by itself evolves around
$M_{2}$.

On a more careful look, we find that the gap $\Delta_{b}$ does change
its structure near $M=M_{2}$ and becomes nodeless. At around the
same $M$, $T_{c}$ is likely non-monotonic and passes through a minimum.

The argument goes as follows. Consider first the pairing confined
to the FS and solve for $T_{c}$. Neglecting $N_{c}$ and $N_{cs}$
in Eq. (\ref{Gap_eqn3}), which are small for all $M$ and $s$, and
solving for the linearized gap equation for $L=\log{\left(\frac{\Lambda}{T_{c}}\right)}$,
we obtain \begin{equation}
1+\frac{U_{3}}{2}\left(sN_{ss}-N\right)L+\left(\frac{U_{3}}{2}\right)^{2}s\left(N_{s}^{2}-NN_{ss}\right)L^{2}=0\label{dec_7_5}\end{equation}

One can verify that $NN_{ss}-N_{s}^{2}$ is positive for all $M>M_{2}$.
For $s\geq1$, when $\Delta_{b}$ has nodes at $M\sim M_{1}$, we
also have $sN_{ss}>N$. In this situation, there is a single solution
of (\ref{dec_7_5}) with $L>0$:

\begin{eqnarray}
L & = & \frac{1}{sU_{3}(NN_{ss}-N_{s}^{2})}\nonumber \\
 &  & \times\left(\sqrt{(sN_{ss}+N)^{2}-4sN_{s}^{2}}+(sN_{ss}-N)\right)\label{dec_7_6}\end{eqnarray}

\begin{figure}[htp]
\includegraphics[width=0.24\textwidth]{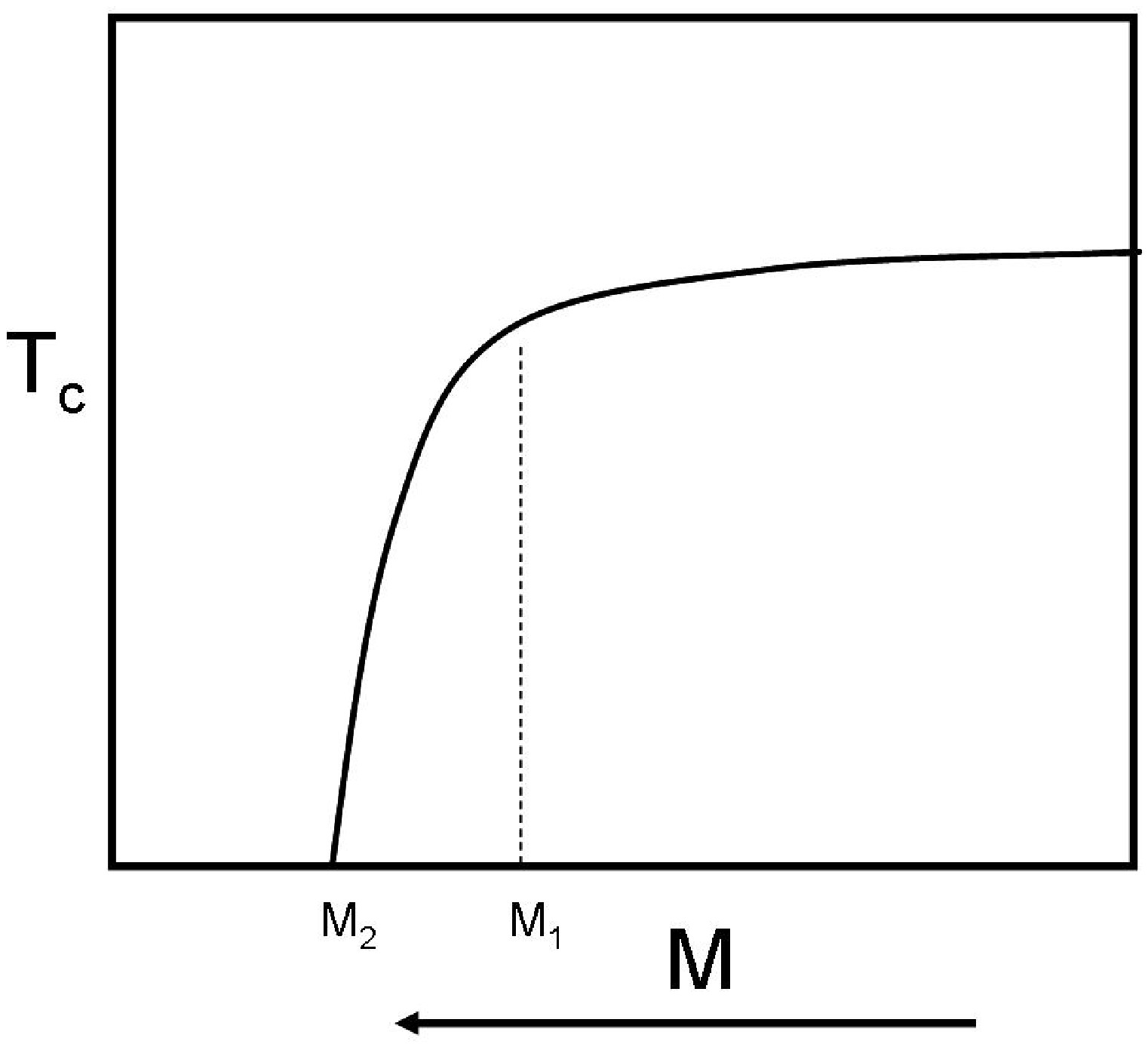}\includegraphics[width=0.22\textwidth]{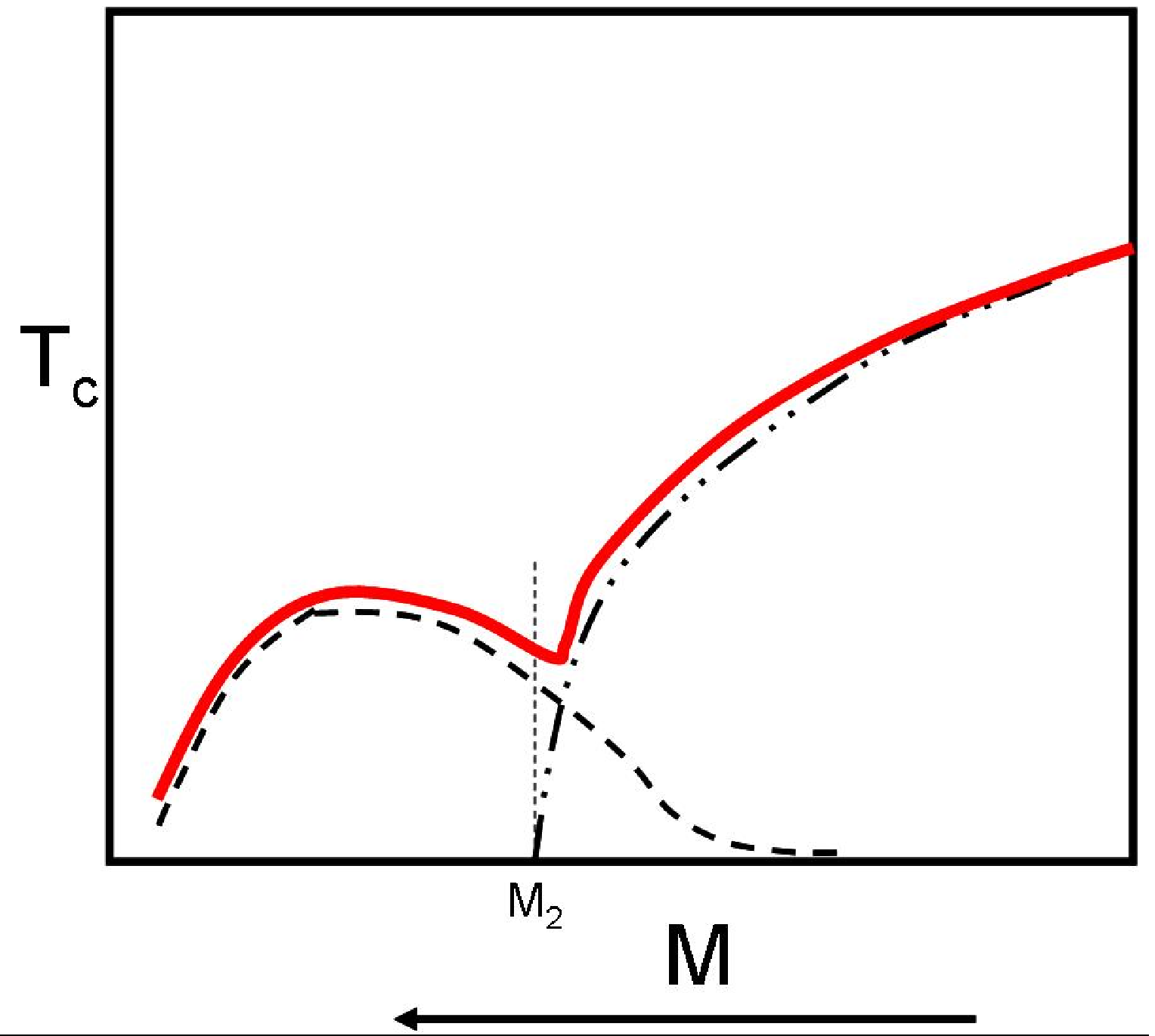}

\caption{\label{fig:schematic} a) Schematic behavior of $T_{c}$ as
a function of $M$, if $T_{c}$ is obtained by restricting to
contributions only from the FSs. This $T_{c}$ drops to zero as
$M\rightarrow M_{2}$. The arrow indicates the direction along
which $M$ increases. b) The actual $T_{c}$ (solid red line) is a
superposition of contributions from the FS and from the gapped
states (dashed lines). This $T_{c}$ initially decreases when $M$
approaches $M_{2}$ but reverses the trend and remains finite at
$M_{2}$ and even in some range of $M>M_{2}$. The dip in $T_{c}$ is
in the region where the contributions from the FS and from the
gapped states become comparable to each other. This  non-monotonic
behavior of $T_{c}$ around $M_{2}$
is captured
by Eq. (\ref{dec_16_1}) (see text). }

\end{figure}

As long as $N_{s}^{2}\ll NN_{ss}$, $L\approx2/U_{3}N$, i.e.
$T_{c}$ does not depend strongly on $M$. However, near $M=M_{2}$,
$\sin2\theta({\bf k})$ on the $b-$pockets tends to a constant
value $c$, see Eq. (\ref{sin_theta_b}), ($c=-0.94$ for our
parameters) and $N_{s}^{2}$ and $NN_{ss}$ both tend to the same
value $c^{2}N^{2}$. As a result, $L$ diverges as
$(2/U_{3}s)(sN_{ss}-N)/(NN_{ss}-N_{s}^{2})$, i.e. $T_{c}$ vanishes
(see Fig \ref{fig:schematic}). One can straightforwardly show that
$L$ diverges at $M=M_{2}$ even if we solve the full $3\times3$
linearized matrix gap equation, without neglecting the terms
$N_{c}$ and $N_{cs}$.

In the same limit $NN_{ss}\approx N_{s}^{2}$, we also have
\begin{equation}
\frac{g_{3}}{g_{1}}\,\sin2\theta_{\mathbf{q}}=-\frac{N_{s}}{N_{ss}}\,
\sin2\theta_{\mathbf{q}}\to-c\frac{-c}{c^{2}}=1.\label{dec_16_3}
\end{equation}
We see that the angle-independent and the $\sin2\theta$ terms in
$\Delta_{b}$ become identical and cancel each other, i.e
$\Delta_{b}$ vanishes (see Eq. (\ref{gaps})). This is consistent
with the vanishing of $T_{c}$.

These results, however, hold only as long as we restrict the
integrals for $N$, $N_{s}$, and $N_{ss}$ to the FS. Once we
include the contributions from the gapped states, $N_{s}^{2}$ and
$NN_{ss}$ no longer tend to the same value. The contributions from
the gapped states contain, instead of $L=\log{\Lambda/T_{c}}$, the
factor $L_{D}=\log{\Lambda/\sqrt{D^{2}+T_{c}^{2}}}$, where $D$ is
the gap in the absence of SC. In a generic weak coupling case,
these contributions would be small in $L_{D}/L$ compared to the
contributions from the FS, but in our case the FS contribution to
$NN_{ss}-N_{s}^{2}$ vanishes at $M_{2}$, and the contribution from
the gapped states become the dominant ones. Similarly,
contributions from the gapped state break the cancelation between
the angle-independent and the $\sin2\theta_{\mathbf{q}}$ terms in
$\Delta_{b}$.

Combining the contributions from the FS and from the gapped states
to $N_{i}$, we have near $M=M_{2}$, \begin{eqnarray}
N & = & N_{0}\left(1+\gamma\frac{L_{D}}{L}\right)\nonumber \\
N_{s} & = & N_{0}\left(c_{s}-\gamma_{s}\frac{L_{D}}{L}\right)\nonumber \\
N_{ss} & = & N_{0}\left(c_{ss}+\gamma_{ss}\frac{L_{D}}{L}\right)\end{eqnarray}
 where $c_{s}\approx-c>0$ and $c_{ss}\approx c^{2}$. One can introduce
$\epsilon\equiv c_{ss}-c_{s}^{2}$ as a measure of deviation from
$M_{2}$, with $\epsilon=0$ at $M_{2}$. The constants $\gamma_{i}$
are all positive, at least if they predominantly come from gapped
$a-$pockets. Since $\sin2\theta_{\mathbf{q}}$ along the
$a-$pockets is quite close to $-1$ (see Fig. \ref{fig:cos-sin}),
we have $\gamma_{s}^{2}=\gamma\gamma_{ss}$ to a good accuracy.
Substituting these forms into the result for $L$, Eq.
(\ref{dec_7_6}), and assuming for simplicity that $s\gg1$, we
obtain for small $\epsilon$

\begin{eqnarray}
 &  & \epsilon{\tilde{L}}^{2}+K_{1}{\tilde{L}}-K_{2}=0\label{dec_16_1}\\
 &  & K_{1}=\left(\gamma c^{2}+\gamma_{ss}+2\gamma_{s}|c|\right){\tilde{L}}_{D}-c^{2}\nonumber \\
 &  & K_{2}=\gamma_{ss}{\tilde{L}}_{D}>0\nonumber \end{eqnarray}

where ${\tilde{L}}=L(N_0 U_{3}/2)$, ${\tilde{L}}_{D}=L_{D}(N_0
U_{3}/2)$. If we set ${\tilde{L}_{D}}=0$, i.e., neglect the
contributions from the gapped states, $K_{2}=0$, $K_{1}=-c^{2}$,
and we obtain ${\tilde{L}}=c^{2}/\epsilon$, as before. If instead
we set $\epsilon=0$, we find that $L$ remains finite and positive,
and to leading order in $\tilde{L}_{D}^{-1}<1$ is given by
\begin{equation}
{\tilde{L}}_{\epsilon=0}\approx\frac{\gamma_{ss}}{\gamma_{ss}+\gamma
c^{2}+2\gamma_{s}|c|}\label{dec_16_2}\end{equation}
The behavior
of ${\tilde{L}}$ at small but finite $\epsilon$ is rather
involved, but its main features can be understood directly from
Eq. (\ref{dec_16_1}). Far from $M_{2}$, $T_{c}$ is not small,
implying that ${\tilde{L}_{D}}$ is small,  i.e and $K_{1}<0$ and
$K_2$ can be neglected. As a result, $\tilde{L}$ increases as
$\tilde{L}\sim1/\epsilon$, leading to a decrease in $T_{c}$ and,
consequently, to an increase in ${\tilde{L}}_{D}$. With increasing
${\tilde{L}}_{D}$,  $K_{1}$ crosses zero and changes sign. At this
point, $K_2 = \mathcal{O}(1)$, and ${\tilde{L}}\sim
K_{2}/\sqrt{\epsilon} \sim 1/\sqrt{\epsilon}$, i.e., it still
increases with decreasing $\epsilon$, but more slowly then before.
Because ${\tilde{L}}_{\epsilon=0}\sim\mathcal{O}(1)$ given by
(\ref{dec_16_2}) is certainly smaller than $1/\sqrt{\epsilon}$,
${\tilde{L}}$ necessary passes through a maximum near the point
where $K_{1}$ changes sign, and then decreases towards
${\tilde{L}}_{\epsilon=0}$. Accordingly, $T_{c}=\Lambda
e^{-2{\tilde{L}}/U_{3}N_{0}}$ passes through a minimum at some
$M\leq M_{2}$, as shown schematically in Fig. \ref{fig:schematic}

Contributions from the gapped states also affect the interplay between
the constant and the $\sin2\theta_{\mathbf{q}}$ terms in $\Delta_{b}$.
if we include these terms, we obtain, near $M=M_{2}$, instead of
Eq. (\ref{dec_16_3}),
\begin{equation}
\frac{g_{3}}{g_{1}}\,\sin2\theta_{\mathbf{q}}=\frac{1+\frac{\gamma_{s}}{c}\frac{L_{D}}{L}}{1+\frac{\gamma_{ss}}{c^{2}}\frac{L_{D}}{L}}\label{dec_16_4}\end{equation}
where, we remind, in our case $\gamma_{s,ss} >0$ and $c=-0.94$. We
see that the numerator gets smaller than one and the denominator
gets larger than one. As a result,
$|\frac{g_{3}}{g_{1}}\,\sin2\theta_{\mathbf{q}}|$ becomes smaller
than one, and the gap $\Delta_{b}$ recovers its nodeless form in
the vicinity of $M_{2}$. Very likely $\Delta_{b}$ remains gapless
also at larger $M$, before $T_{c}$ finally vanishes.

We also note that our consideration is self-consistent in the
sense that the transformation from nodal to nodeless gap and
non-vanishing of $T_{c}$ at $M=M_{2}$, where $\epsilon=0$, are
consistent with each other. Indeed, once the gap $\Delta_{b}$
becomes nodeless, the pairing problem at $M\geq M_{2}$ is
qualitatively similar to the one for fermions with the dispersion
\begin{equation}
\varepsilon_{gapped}(k)=\sqrt{D^{2}+\left[\mathbf{v}_{F}\cdot\left(\mathbf{k}-\mathbf{k}_{F}\right)\right]^{2}},\label{e1}\end{equation}
where $D$ is the pre-existing gap in the excitation spectrum. For
such model, $T_{c}$ is non-zero at $D=0$ (equivalent of $M=M_{2}$)
and extends into the region where $D\neq0$: \begin{equation}
T_{c}(D)=T_{c}(0)F\left[\frac{D}{T_{c}(0)}\right]\label{dec_7_9}\end{equation}
where $F(x)$ is a decreasing function of $x$ subject to
$F(x\rightarrow0)=1-x^{2}$ and $F(x)=-1/\ln\left(1.76-x\right)$
near the critical $x=1.76$.

The outcome of this analysis is that nodal SC gap exists only in relatively
narrow range of $M$, from $M\leq M_{1}$ to $M\leq M_{2}$. At larger
$M$ the gap is again nodeless, and, furthermore, for $M>M_{2}$ all
low-energy states are gapped already above $T_{c}$. The SC transition
temperature $T_{c}$ is non-zero at $M_{2}$ and decreases into $M>M_{2}$
region. Before that, $T_{c}$ has a minimum roughly where the gap
changes from nodal to nodeless.

%SM commented
%The minimum in $T_{c}$ in the regime where the gap has nodes has
%been recently observed in high-accuracy measurements of $T_{c}$ in
%underdoped (Ba$_{1-x}$K$_{x}$)Fe$_{2}$As$_{2}$ (Ref.
%\onlinecite{ruslan_private}). Our result is in full agreement with
%this data.

\subsection{Role of the non-reconstructed pockets}

\label{sec:spectator}

So far, we have restricted our consideration of the pairing problem
to one hole and one electron pocket reconstructed by SDW. There are
additional pockets which do not participate in the SDW state \cite{eremin_09}.
In this subsection we analyze to what extent these additional pockets
affect our results. Specifically, we add another elliptical electron
pocket centered at $\left(\pi,0\right)$, denote fermions near this
pocket by $d_{\mathbf{k}}$, and solve the set of coupled linearized
gap equations for $\Delta_{a}(\mathbf{q})$, $\Delta_{b}({\bf \mathbf{q}})$,
and $\Delta_{d}({\bf \mathbf{q}})$. The interactions involving fermions
near $\left(0,0\right)$ and near the electron pocket at $(\pi,0)$
are the same as the interactions between $\left(0,0\right)$ and $(0,\pi)$,
i.e., the couplings are the same $U_{1}$, $U_{2}$, $U_{3}$, and
$U_{4}$. As before, we consider the model with only $U_{3}$ and
$U_{1}=sU_{3}$. We also neglect direct interaction between electron
pockets.

The effective pairing interactions in the $a-b-d$ space are obtained
by dressing the interactions by coherence factors associated with
the transformation from $c$ and $f$ to $a$ and $b$ operators for
the pockets at $\left(0,0\right)$ and at $(0,\pi)$. Since the expressions
for the vertices are long, we refrain from presenting them. Yet, it
is straightforward to obtain and solve the set of linearized gap equations
for $\Delta_{a}({\bf \mathbf{q}})$, $\Delta_{b}({\bf \mathbf{q}})$,
and $\Delta_{d}(\mathbf{q})$. It is essential that the \textquotedbl{}dressing\textquotedbl{}
does not involve $d-$fermions, hence the interaction has no dependence
on the angle along the $d-$FS. As a consequence, $\Delta_{d}({\bf \mathbf{q}})$
remains angle-independent. This constant $\Delta_{d}$, however, affects
the interplay between the angle-independent and the $\cos2\theta_{\mathbf{k}}$
and $\sin2\theta_{\mathbf{k}}$ terms for $\Delta_{a}(\mathbf{q})$
and $\Delta_{b}(\mathbf{q})$.

\begin{figure*}[htp]
 \includegraphics[width=0.3\textwidth]{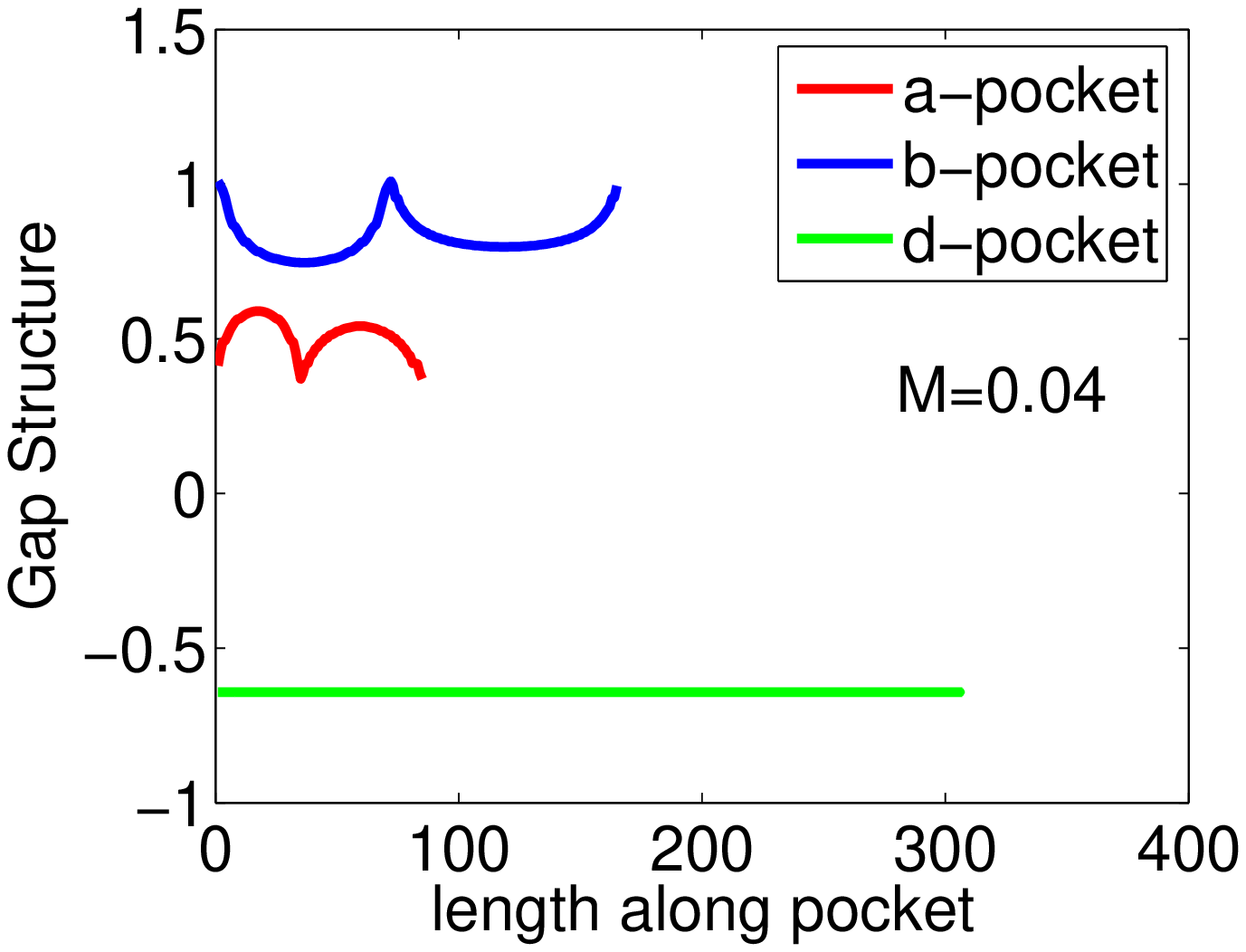} \includegraphics[width=0.3\textwidth]{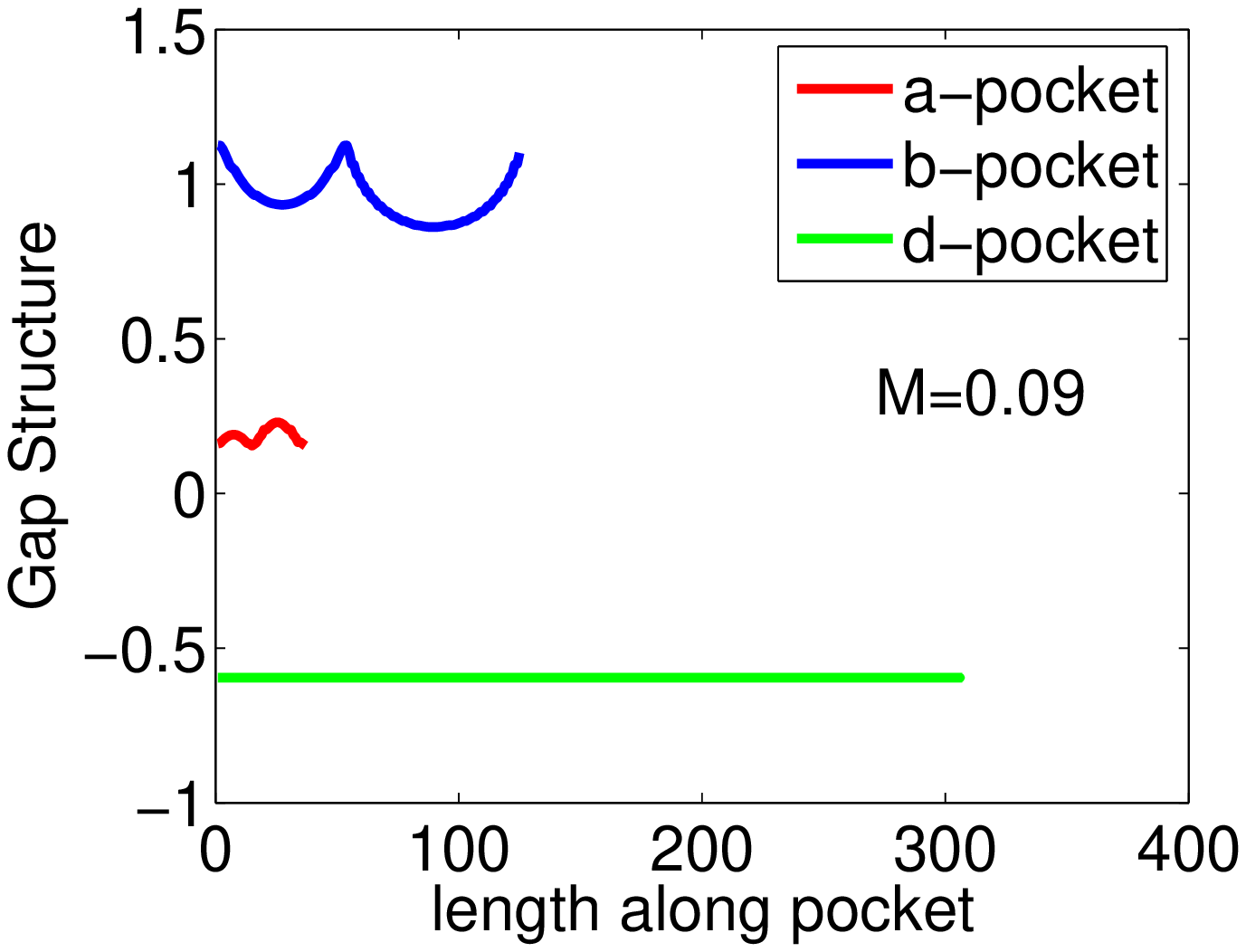}
\includegraphics[width=0.3\textwidth]{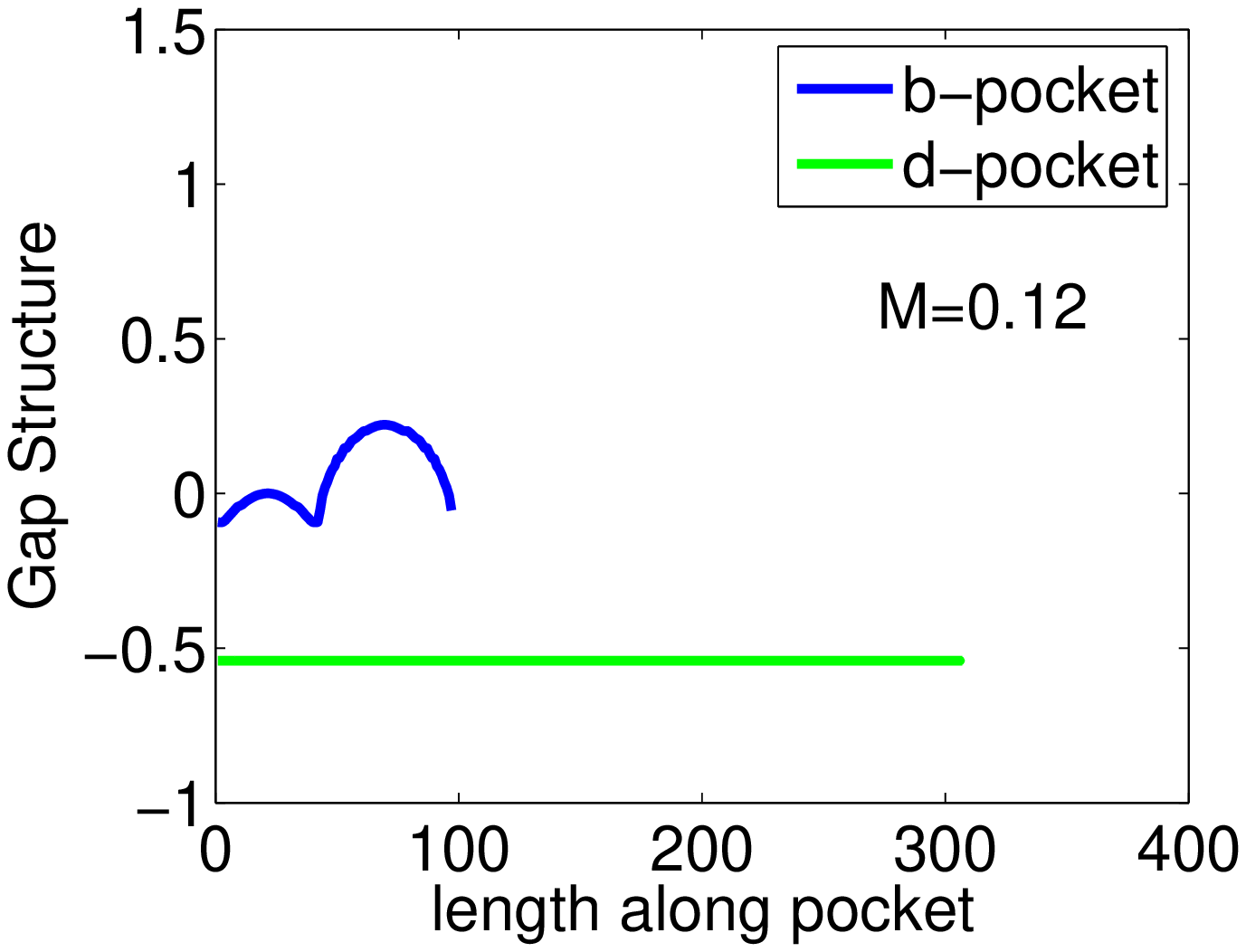}

\caption{\label{fig:gap_3poc} The gap structure for $s=5$ and the same values
of $M$ as used in Fig. \ref{fig:FS_gaps} for the case when we include
an additional, spectator pocket at $(\pi,0)$ which does not participate
in the FS reconstruction. A comparison with Fig. \ref{fig:FS_gaps}
shows that the gap structure is almost unaffected by the presence
of the spectator pocket.}

\end{figure*}

In Fig. \ref{fig:gap_3poc}, we show the gap structure obtained from
the solution of the pairing problem for $s=5$ and the same three
values of $M$ as in Fig. \ref{fig:FS_gaps}. Comparing Figs. \ref{fig:FS_gaps}
and Fig. \ref{fig:gap_3poc}, we see that the non-reconstructed $d$-pocket
has only a minor effect on the $a-b$ gap structure -- the gap on
the $b$ pocket still becomes more angle-dependent with increasing
$M$ and develops accidental nodes at $M\geq M_{1}$. In view of this
result, it is very likely that the physics of gap variation with $M$
is fully captured already within the two-band model of one hole and
one electron pocket. %

\section{Effect of an external magnetic field}

\label{sec:magnetic field}

We now consider how the SC gap structure is affected by an external
magnetic field $\mathbf{H}$. The specific goal is to verify whether
the field increases or reduces the tendency towards the development
of nodes in $\Delta_{b}$. This issue is related to experiments on
(Ba$_{1-x}$K$_{x}$)Fe$_{2}$As$_{2}$, particularly to recent measurements
of thermal conductivity in a field~\cite{JPR}.

We direct the field along $z$ and include into the Hamiltonian the
Zeeman coupling between the field and the $z-$component of the spin
of a fermion $S_{z}=(1/2)c_{\alpha}^{\dagger}\sigma_{\alpha\beta}^{z}c_{\beta}$.
For an isotropic system, the SDW vector $\mathbf{M}$ is oriented
transverse to $\mathbf{H}$ and for definiteness we direct it along
$x$. At a finite $H$, the system also develops a non-zero uniform
magnetization, which rotates a spin at a given site towards the field
and creates a canted two-sublattice structure, with orthogonal antiferromagnetic
SDW component ${\bf M}=M_{x}$ and ferromagnetic component along $z$.

The quadratic part of the Hamiltonian in the presence of the field
is

\begin{eqnarray}
H & = & \sum_{\mathbf{k},\sigma}\left[\left(\varepsilon_{\mathbf{k}}^{c}+\sigma h\right)c_{\mathbf{k}\sigma}^{\dagger}c_{\mathbf{k}\sigma}+\left(\varepsilon_{\mathbf{k}+\mathbf{Q}}^{f}+\sigma h\right)f_{\mathbf{k}+\mathbf{Q}\sigma}^{\dagger}f_{\mathbf{k}+\mathbf{Q}\sigma}^{}\right]\nonumber \\
 & + & \sum_{\mathbf{k}}M\left(c_{\mathbf{k}\uparrow}^{\dagger}f_{\mathbf{k}+\mathbf{Q}\downarrow}^{}+c_{\mathbf{k}\downarrow}^{\dagger}f_{\mathbf{k}+\mathbf{Q}\uparrow}^{}+\mathrm{h.c.}\right)\end{eqnarray}
 where $h=\mu_{B}H$. We will measure $h$ in units of $2\mu$, as
with other observables.

The transformation to the new operators is now given by \begin{eqnarray}
\left(\begin{array}{c}
\sigma a_{\mathbf{k}\sigma}\\
b_{\mathbf{k}\sigma}\end{array}\right) & = & \left(\begin{array}{cc}
-\sigma\cos\theta_{\mathbf{k},\sigma} & \sin\theta_{\mathbf{k},\sigma}\\
-\sin\theta_{\mathbf{k},\sigma} & -\sigma\cos\theta_{\mathbf{k},\sigma}\end{array}\right)\left(\begin{array}{c}
c_{\mathbf{k}\sigma}\\
f_{\mathbf{k}+\mathbf{Q}\sigma}\end{array}\right)\end{eqnarray}
 where

\begin{eqnarray}
\cos\theta_{\mathbf{k},\sigma} & = & \frac{M}{\sqrt{M^{2}+(E_{\sigma}^{-})^{2}}}\nonumber \\
\sin\theta_{\mathbf{k},\sigma} & = & \frac{E_{\sigma}^{-}}{\sqrt{M^{2}+(E_{\sigma}^{-})^{2}}}\label{eq:}\end{eqnarray}
 and \begin{equation}
E_{\sigma}^{-}=\left(\frac{\varepsilon^{c}-\varepsilon^{f}}{2}+\sigma h\right)-\sqrt{\left(\frac{\varepsilon^{c}-\varepsilon^{f}}{2}+\sigma h\right)^{2}+M^{2}}\end{equation}

Substituting the transformation into $H_{int}$ in Eq. (\ref{1})
and taking care of the fact that $\cos\theta_{k,\uparrow}$ and $\cos\theta_{k,\downarrow}$
are now different, we obtain Eq. (\ref{H_pair}) but with new $U_{\mathbf{q},\mathbf{k}}^{aa}$,
$U_{\mathbf{q},\mathbf{k}}^{bb}$ and $U_{\mathbf{q},\mathbf{k}}^{ab}$
given by \begin{widetext} \begin{eqnarray*}
U_{\mathbf{q},\mathbf{k}}^{aa} & = & U_{\mathbf{q},\mathbf{k}}^{bb}=-\frac{U_{3}}{2}\left(C_{\mathbf{q}\uparrow}C_{-\mathbf{q}\downarrow}S_{\mathbf{k}\uparrow}S_{-\mathbf{k}\downarrow}+S_{\mathbf{q}\uparrow}S_{-\mathbf{q}\downarrow}C_{\mathbf{k}\uparrow}C_{-\mathbf{k}\downarrow}\right)+U_{1}C_{\mathbf{q}\uparrow}S_{-\mathbf{q}\downarrow}C_{\mathbf{k}\uparrow}S_{-\mathbf{k}\downarrow}\\
U_{\mathbf{q},\mathbf{k}}^{ab} & = & U_{\mathbf{q},\mathbf{k}}^{ba}=-\frac{U_{3}}{2}\left(C_{\mathbf{q}\uparrow}C_{-\mathbf{q}\downarrow}C_{\mathbf{k}\uparrow}C_{-\mathbf{k}\downarrow}+S_{\mathbf{q}\uparrow}S_{-\mathbf{q}\downarrow}S_{\mathbf{k}\uparrow}S_{-\mathbf{k}\downarrow}\right)-U_{1}C_{\mathbf{q}\uparrow}S_{-\mathbf{q}\downarrow}C_{\mathbf{k}\uparrow}S_{-\mathbf{k}\downarrow}\end{eqnarray*}
 \end{widetext} where $C_{\mathbf{k},\sigma}\equiv\cos\theta_{\mathbf{k,}\sigma}$
and $S_{\mathbf{k},\sigma}\equiv\sin\theta_{\mathbf{k},\sigma}$.
The gap structure consistent with these $U_{\mathbf{q},\mathbf{k}}^{ij}$
is of the form \begin{eqnarray*}
\Delta_{a}(\mathbf{q}) & = & g_{1}C_{\mathbf{q}\uparrow}C_{-\mathbf{q}\downarrow}+g_{2}S_{\mathbf{q}\uparrow}S_{-\mathbf{q}\downarrow}\\
 &  & +g_{3}C_{\mathbf{q}\uparrow}S_{-\mathbf{q}\downarrow}+g_{4}S_{\mathbf{q}\uparrow}C_{-\mathbf{q}\downarrow}\\
\Delta_{b}(\mathbf{q}) & = & g_{1}C_{\mathbf{q}\uparrow}C_{-\mathbf{q}\downarrow}+g_{2}S_{\mathbf{q}\uparrow}S_{-\mathbf{q}\downarrow}\\
 &  & -g_{3}C_{\mathbf{q}\uparrow}S_{-\mathbf{q}\downarrow}-g_{4}S_{\mathbf{q}\uparrow}C_{-\mathbf{q}\downarrow}\end{eqnarray*}

\begin{figure}[htp]
\includegraphics[width=0.23\textwidth]{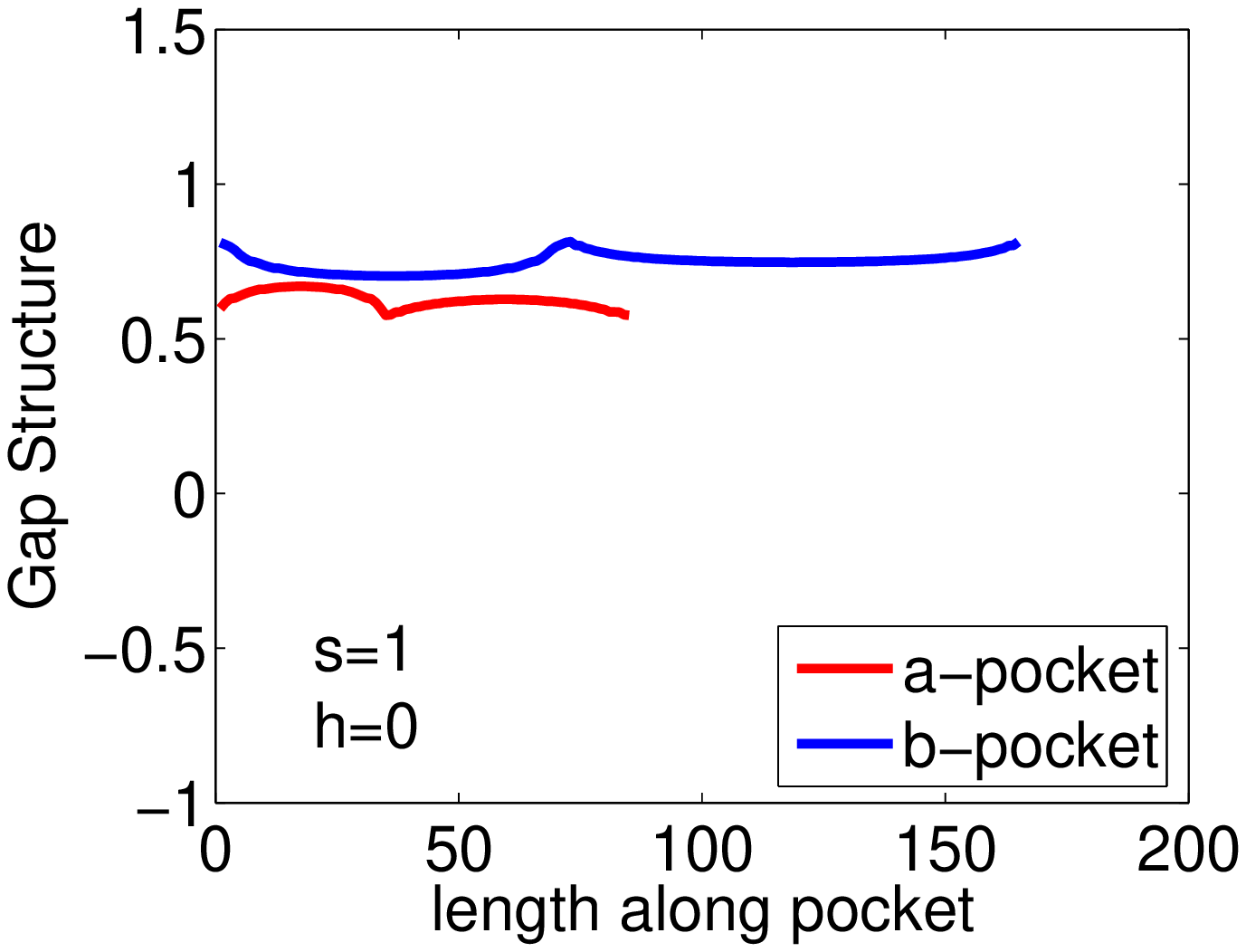} \includegraphics[width=0.23\textwidth]{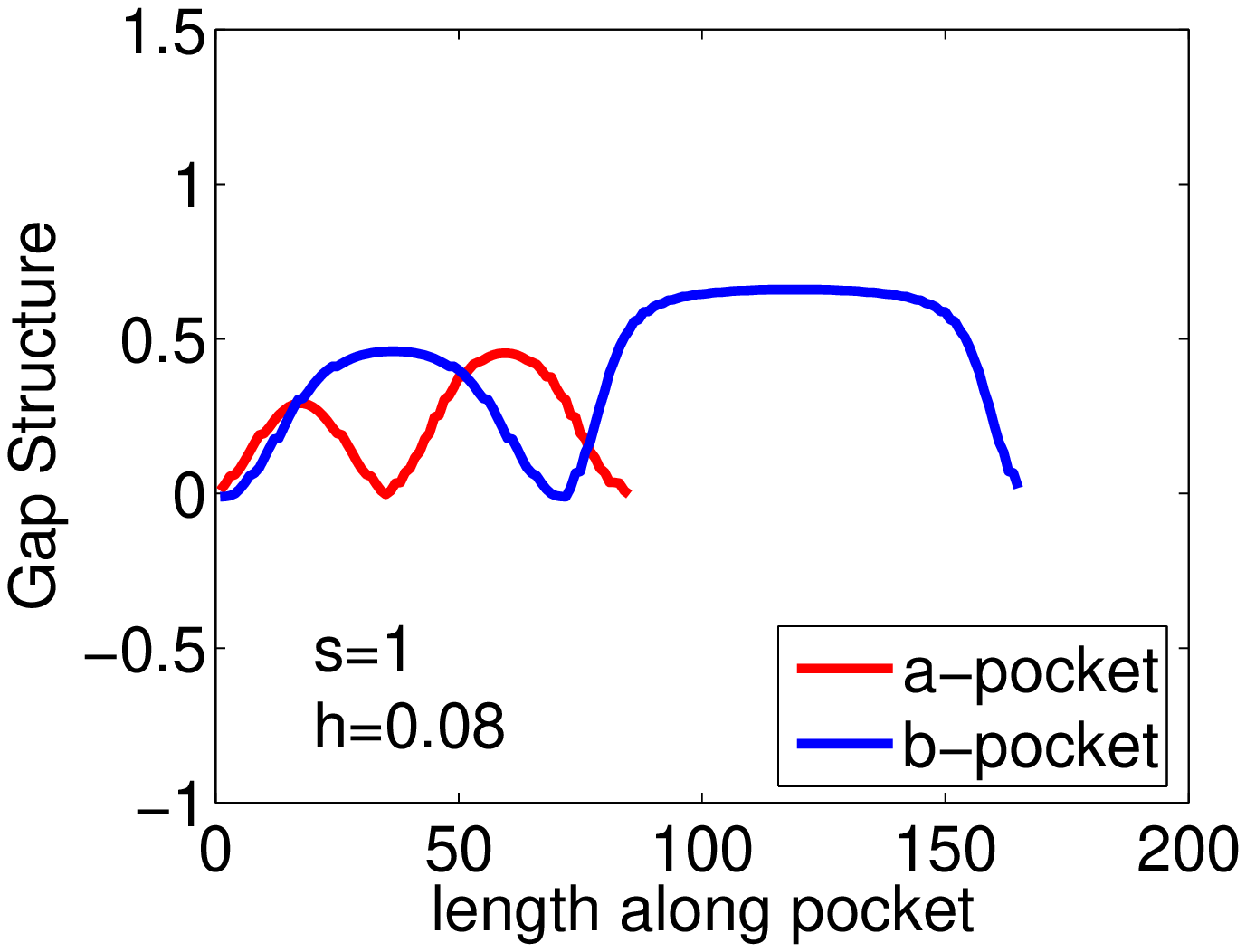}

\caption{\label{fig:gap_mag_field} The gap structure for $h=0$ (left panel)
and $h=0.08$ (right panel) for $M=0.04$ and $s=1$. The application
of the field increases angle variations of the gaps and leads to the
appearance of nodes at a smaller $M$ (and a smaller $s$) than without
a field.}

\end{figure}

We constructed the set of coupled linearized gap equations for $g_{i}$
by standard means and solved them for various $M$, $s=U_{1}/U_{3}$
and $h$. We found that the field enhances the angle variation of
the gaps $\Delta_{a}$ and $\Delta_{b}$, such that nodes appear at
smaller $M$ and smaller $s$. To illustrate this, in Fig. \ref{fig:gap_mag_field}
we compare the gap structure for a given $M=0.04$ and $s=1$ (same
as in the upper left panel in Fig. \ref{fig:FS_gaps}) without a field
and with a field. We clearly see that the gap variations along the
reconstructed FSs grow with the field, and for large enough field
nodes appear well before $M$ reaches the value at which the $a-$pockets
disappear. The implication is that the range where the gap has nodes
widens up with the application of a field and, in particular, extends
to smaller $M$ (larger dopings), where without a field the system
was a superconductor with a nodeless gap.

We emphasize that the extension of the nodal region to smaller $M$
is a different effect than the field-induced changes in $M_{1}$ and
$M_{2}$. These last changes are due to the modification of the quasiparticle
dispersion, which in the presence of the field becomes 4-band dispersions
and takes the form \begin{equation}
E_{\mathbf{k}}^{a,b}=\frac{\varepsilon^{c}+\varepsilon^{f}}{2}\pm\sqrt{\left(\frac{\varepsilon^{c}-\varepsilon^{f}}{2}\pm h\right)^{2}+M^{2}}\end{equation}

The two $a$-pockets are split by $\pm h$ into $a^{+}$ and $a^{-}$
pockets, and the two $b-$pockets are split by $\pm h$ into $b^{+}$
and $b^{-}$ pockets.

The presence of $\pm h$ under the square root splits $M_{1}$ and
$M_{2}$ into $M_{1}^{+},M_{1}^{-}$ and $M_{2}^{+},M_{2}^{-}$,
where $M_{1,2}^{+}=M_{1,2}(1+2h)$ and $M_{1,2}^{-}=M_{1,2}(1-2h)$,
with $M_{1}$ and $M_{2}$ given by Eqs \ref{M_1} and \ref{M_2} (we
recall that $h$ is measured in units of $2\mu$). For $h=0.08$ used
in Fig. \ref{fig:gap_mag_field} we get $M_{1}^{+}=0.118,\,
M_{1}^{-}=0.086$ instead of $M_{1}=0.102$, and $M_{2}^{+}=0.205,\,
M_{2}^{-}=0.149$ instead of $M_{2}=0.177$. Clearly, the
field-induced changes in the form of reconstructed FSs are small
compared to the changes of the SC gap structure associated with
the field-induced change of the pairing interaction.
%SM commented
The small effect of the field on the reconstructed FS
is in agreement with previous works on the cuprates \cite{Subir}.
%and also on experimental results in the pnictides, which find no change
%in $T_{N}$ upon the application of an external magnetic field \cite{JPR_comm}.

\section{Discussion and concluding remarks}

\label{sec:conclusion}

We showed that the superconducting gap of an $s^{+-}$ SC in the co-existence
phase with an SDW order acquires additional angular dependence compared
to the case when SDW order is not present. In particular, an isotropic
pairing interaction becomes angle-dependent inside the SDW state.
At small SDW order parameter $M$, or when the density-density interaction
between the hole and electron pockets is small compared to pair hopping
interaction term, this extra dependence is weak, in agreement with
previous studies\cite{parker_09,rafael}. However, when density-density
and pair hopping interactions are comparable and $M$ is large enough
to gap out one out of two pairs of reconstructed FS pockets, the angular
dependence gets quite strong and the SC gap on the remaining FS banana-like
pocket develops accidental nodes near the tips of the bananas. At
even larger $M$, the system bounces back into the fully gapped SC
state which extends into the regime where large enough $M$ gaps out
the remaining pair of FS pockets. The SC transition temperature $T_{c}$
has a dip near the point where nodal SC state becomes nodeless. In
the presence of a magnetic field, the width of the nodal region expands,
and, in particular, the nodes appear at smaller $M$ (larger doping).

Our results offer a consistent explanation for the recent experimental
observations\cite{JPR} in underdoped (Ba$_{1-x}$K$_{x}$)Fe$_{2}$As$_{2}$.
Performing in-plane thermal conductivity measurements, Reid \emph{et
al.} found a small range of doping in the SC-SDW coexistence region
where the ratio $\kappa/T$ is finite at $T=0$, i.e., $\kappa$ is
linear in $T$ at small $T$. This linearity is generally viewed as
a strong indication for the presence of the nodes in the SC gap. At
smaller and larger dopings Reid \emph{et al.} found that $\kappa/T$
vanishes at $T=0$, as it is expected for a fully gapped superconductor.
They also observed that the doping range where $\kappa/T$ is finite
at $T=0$ expands upon the application of an external magnetic field.

In terms of our model, the nodeless-nodal-nodeless transition observed
by Reid \emph{et al.} can be attributed to the following sequence
of events: at the optimal doping, which roughly coincides with the
onset of the coexistence range, the SC gaps in (Ba$_{1-x}$K$_{x}$)Fe$_{2}$As$_{2}$
are almost isotropic, and the excitation spectrum is fully gapped.
Decreasing $x$ and moving to the underdoped region in which SC and
SDW start to coexist is equivalent to making $M$ nonzero in our model.
In this situation, the gaps become anisotropic, but remain nodeless
over some range of $M$. Moving deeper inside the coexistence region
is equivalent to making $M$ larger. Eventually, $M$ becomes large
enough such that one of the reconstructed pairs of pockets vanishes.
In this situation, we found (for $s\geq1$) that the gap on the remaining
FS pockets develops nodes, what leads to a finite $\kappa/T$ at $T=0$.
As the doping level decreases even further, $M$ continues to increase
and the gap structure bounces back to nodeless, since for such pairing
state $T_{c}$ remains non-zero even when the remaining pair of reconstructed
pockets disappears. The quasiparticle spectrum becomes fully gapped
again, and $\kappa/T$ vanishes at $T=0$.

In this explanation the nodeless-nodal transition is roughly
associated with the vanishing of one pair of reconstructed FS
pockets. An SDW-driven electronic transition has been observed by
ARPES as well as by transport measurements in electron-doped
Ba(Fe$_{1-x}$Co$_{x}$)$_{2}$As$_{2}$~\cite{Lifshitz_transition}.
In that case, however, the reconstructed pocket that disappears is
a much smaller hole-like pocket associated with additional details
of the band dispersion not captured by our simplified two-band
model. It would be interesting to verify experimentally whether
the transition from four to two pockets takes place in the
hole-doped (Ba$_{1-x}$K$_{x}$)Fe$_{2}$As$_{2}$ materials, and what
is their relationship to the SC gap structure. Such transitions
should have distinct signatures not only in the band dispersions,
but also in transport coefficients. Furthermore, as discussed in
Refs. \onlinecite{scaling_nodes1,scaling_nodes2}, the onset of
accidental nodes should also affect the low-temperature behavior
of several thermodynamic quantities, giving rise to peculiar
scaling relations. Another result of our analysis is the
observation of a dip in the doping dependence of $T_{c}$ near the
doping where the system undergoes the nodal-nodeless transition.

%SM commented
%The dip in $T_{c}$ at the boundary
%of the nodal region has been recently observed~\cite{ruslan_private}
%in high-accuracy measurements of $T_{c}$ in (Ba$_{1-x}$K$_{x}$)Fe$_{2}$As$_{2}$
%at various dopings in the coexistence regime. Our result is fully
%consistent with the data.

Our model indeed does not include all aspects of the physics of
the co-existence state. In particular, doping with holes or
electrons  changes the chemical potential, which was kept constant
in our calculation. The feedback from this change affects the
values $M_{1}$ and $M_{2}$ at which reconstructed pockets
disappear leaving a possibility that reconstructed FSs can be
present even for zero doping.  The angular dependence of the
magnetic interaction may also preserve FS pockets even for large
$M$ ~\cite{vishwanath}. Still, we believe that the physics
described by our model is quite genetic and should hold for more
realistic models.

Finally, we point out that the disappearance of two out of four pockets
is not the only mechanism that can give rise to nodes in the coexistence
state. When the exchange interaction between unreconstructed hole
and electron pockets is strong enough, it can induce nodes even in
the absence of any dramatic change of the reconstructed FSs. Conversely,
there is also a region in parameter space (small $s$) in which the
coexistence with SDW does not generates nodes in the SC gap. \bigskip{}

We are thankful to L. Taillefer, J.-P. Reid, R. Prozorov, M.
Tanatar, J. Schmalian,  I. Eremin, J. Knolle, for useful
discussions and for sharing unpublished results with us. The work
was supported by NSF-DMR-0906953 (S. M and A.V.C) and by the NSF
Partnerships for International Research and Education (PIRE)
program (R.M.F.). A.V.C gratefully acknowledges partial support
from Humboldt foundation.


\begin{thebibliography}{10}
\bibitem{BFCA_1} K. Terashima, Y. Sekiba, J. H. Bowen, K. Nakayama,
T. Kawahara, T. Sato, P. Richard, Y.-M. Xu, L. J. Li, G. H. Cao, Z.-A.
Xu, H. Ding and T. Takahashi, Proc. Natl. Acad. Sci. USA \textbf{106},
7330 (2009)

\bibitem{BFCA_2} K. Gofryk, A. S. Sefat, M. A. McGuire, B. C. Sales,
D. Mandrus, J. D. Thompson, E. D. Bauer, and F. Ronning, Phys. Rev.
B. \textbf{81}, 184518(2010).

\bibitem{BFCA_3} R. T. Gordon, N. Ni, C. Martin, M. A. Tanatar, M.
D. Vannette, H. Kim, G. D. Samolyuk, J. Schmalian, S. Nandi, A. Kreyssig,
A. I. Goldman, J. Q. Yan, S. L. Bud'ko, P. C. Canfield, and R. Prozorov,
Phys. Rev. Lett. \textbf{102}, 127004 (2009).

\bibitem{BKFA_1} Y-M. Xu, Y-B. Huang, X-Y. Cui, E. Razzoli, M. Radovic,
M. Shi, G-F. Chen, P. Zheng, N-L. Wang, C-L. Zhang, P-C. Dai, J-P.
Hu, Z. Wang, and H. Ding, Nat. Phys. \textbf{7}, 198 (2011).

\bibitem{BKFA_2} X. G. Luo, M. A. Tanatar, J.-Ph. Reid, H. Shakeripour,
N. Doiron-Leyraud, N. Ni, S. L. Bud'ko, P. C. Canfield, Huiqian Luo,
Zhaosheng Wang, Hai-Hu Wen, R. Prozorov, and Louis Taillefer, Phys.
Rev. B. \textbf{80}, 140503 (2009)

\bibitem{BKFA_3} C. Martin, R. T. Gordon, M. A. Tanatar, H. Kim,
N. Ni, S. L. Bud'ko, P. C. Canfield, H. Luo, H. H. Wen, Z. Wang, A.
B. Vorontsov, V. G. Kogan, and R. Prozorov, Phys. Rev. B. \textbf{80},
020501(2009).

\bibitem{BFAP_1} T. Shimojima, F. Sakaguchi, K. Ishizaka, Y. Ishida,
T. Kiss, M. Okawa, T. Togashi, C.-T. Chen, S. Watanabe, M. Arita,
K. Shimada, H. Namatame, M. Taniguchi, K. Ohgushi, S. Kasahara, T.
Terashima, T. Shibauchi, Y. Matsuda, A. Chainani, and S. Shin, Science
\textbf{332}, 564 (2011).

\bibitem{BFAP_2} K. Hashimoto, M. Yamashita, S. Kasahara, Y. Senshu1,
N. Nakata, S. Tonegawa, K. Ikada, A. Serafin, A. Carrington, T. Terashima,
H. Ikeda, T. Shibauchi, and Y. Matsuda, Phys. Rev. B. \textbf{81},
220501(2010).

\bibitem{coexist_1} D. K. Pratt, W. Tian, A. Kreyssig, J. L. Zarestky,
S. Nandi, N. Ni, S. L. Bud'ko, P. C. Canfield, A. I. Goldman, and
R. J. McQueeney, Phys. Rev. Lett. \textbf{103}, 087001 (2009).

\bibitem{coexist_2} S. Avci, O. Chmaissem, E. A. Goremychkin, S.
Rosenkranz, J.-P. Castellan, D. Y. Chung, I. S. Todorov, J. A. Schlueter,
H. Claus, M. G. Kanatzidis, A. Daoud-Aladine, D. Khalyavin, and R.
Osborn Phys. Rev. B. \textbf{83}, 172503(2011).

\bibitem{coexist_3} M.-H. Julien, H. Mayaffre, M. Horvatic, C. Berthier,
X. D. Zhang, W. Wu, G. F. Chen, N. L. Wang and J. L. Luo, Eur. Phys.
Lett. \textbf{87} 37001 (2009).

\bibitem{coexist_4} E. Wiesenmayer, H. Luetkens, G. Pascua, R. Khasanov,
A. Amato, H. Potts, B. Banusch, H.-H. Klauss, and D. Johrendt, Phys.
Rev. Lett. \textbf{107}, 237001 (2011).

\bibitem{parker_09} D. Parker, M. G. Vavilov, A. V. Chubukov, and
I. I. Mazin, Phys. Rev. B \textbf{80}, 100508 (2009).

\bibitem{vvc_10} A. B. Vorontsov, M. G. Vavilov, and A. V. Chubukov,
Phys. Rev. B 81, 174538 (2010).

\bibitem{rafael} R. M. Fernandes and J. Schmalian , Phys. Rev. B
\textbf{82}, 014521 (2010).

\bibitem{fernandes_optics} R. M. Fernandes and J. Schmalian , Phys.
Rev. B \textbf{82}, 014520 (2010).

\bibitem{eremin_09} I. Eremin and A. V. Chubukov, Phys. Rev. B 81,
024511 (2010); J. Schmiedt, P. M. R. Brydon, and C. Timm, arXiv:1108.5296.

\bibitem{fernandes_11} R. Fernandes \textit{et al}, arXiv:1110.1893.

\bibitem{Andrey} A.V. Chubukov, Physica C \textbf{469}, 640 (2009).

\bibitem{comm} The actual situation may be more complex as at least
in some orbital models SDW gap turns out to be angular dependent and
vanishes along particular directions~\cite{vishwanath}. Near these
directions FS survives even when SDW order parameter is large.

\bibitem{vishwanath} Y. Ran, F. Wang, H. Zhai, A. Vishwanath, and
D.-H. Lee, Phys. Rev. B \textbf{79}, 014505 (2009).

\bibitem{JPR} J.-Ph. Reid, M. A. Tanatar, X. G. Luo, H. Shakeripour,
S. René de Cotret, A. Juneau-Fecteau, N. Doiron-Leyraud, J. Chang,
B. Shen, H.-H. Wen, H. Kim, R. Prozorov, and Louis Taillefer,
arXiv:1105.2232 (2011);   R. Prozorov and M. Tartar, private
communication; J.-Ph. Reid and Louis Taillefer, APS March Meeting
Abs: A22.00004, (2012)

\bibitem{ruslan_private} R. Prozorov and M. Tartar, private communication.

\bibitem{maxim} M. G. Vavilov, A. V. Chubukov, and A. B. Vorontsov,
Phys. Rev. B 84, 140502 (2011); M. G. Vavilov and A. V. Chubukov,
arXiv:1110.0972

\bibitem{Subir} Y. Zhang, E. Demler, and S. Sachdev, Phys. Rev. B.
\textbf{66}, 094501 (2002)

%\bibitem{JPR_comm} J-P Reid, L. Taillefer, private communication.

\bibitem{Lifshitz_transition} C. Liu \emph{et al.,} Nature Phys.
\textbf{6}, 419 (2010).

\bibitem{scaling_nodes1} Valentin Stanev, Boian S. Alexandrov, Predrag
Nikolic, and Zlatko Tesanovic, Phys. Rev. B \textbf{84}, 014505 (2011).

\bibitem{scaling_nodes2} R. M. Fernandes and J. Schmalian, Phys.
Rev. B \textbf{84}, 012505 (2011).
\end{thebibliography}
\end{document}